\documentclass[default]{sn-jnl}

\usepackage{graphicx}%
\usepackage{multirow}%
\usepackage{amsmath,amssymb,amsfonts}%
\usepackage{amsthm}%
\usepackage{mathrsfs}%
\usepackage[title]{appendix}%
\usepackage{xcolor}%
\usepackage{textcomp}%
\usepackage{manyfoot}%
\usepackage{booktabs}%
\usepackage{algorithm}%
\usepackage{algorithmicx}%
\usepackage{algpseudocode}%
\usepackage{listings}%

\usepackage{subcaption}
\usepackage{booktabs}
\PassOptionsToPackage{table}{xcolor}
\usepackage{colortbl}
\usepackage{soul}
\usepackage{booktabs, makecell}
\usepackage{tabularx}
\usepackage[T1]{fontenc}
\usepackage{etoolbox}
\usepackage{graphicx}

\usepackage{xr}
\definecolor{aliceblue}{rgb}{0.94, 0.97, 1.0}
\definecolor{bubblegum}{rgb}{0.99, 0.76, 0.8}

\begin{document}


\newbool{showComments}
\booltrue{showComments}
\ifbool{showComments}{%
\newcommand{\forSagar}[1]{\sethlcolor{orange}\hl{[TODO Sagar: #1]}}
\newcommand{\forSanja}[1]{\sethlcolor{aliceblue}\hl{[TODO Sanja: #1]}}
\newcommand{\forAdam}[1]{\sethlcolor{yellow}\hl{[TODO Adam: #1]}}
\newcommand{\forStephen}[1]{\sethlcolor{bubblegum}\hl{[TODO/FROM Stephen: #1]}}
\newcommand{\todo}[1]
{\sethlcolor{red}\hl{[TODO: #1]}}
\newcommand\sss[1]{\textcolor{black}{#1}}
}
{
\newcommand{\forSagar}[1]{}
\newcommand{\forSanja}[1]{}
\newcommand{\forAdam}[1]{}
\newcommand{\forStephen}[1]{}
\newcommand{\todo}[1]{}
\newcommand\sss[1]{\textcolor{black}{#1}}
}


\title{Vitamin N: Benefits of Different Forms of Public Greenery for Urban Health} 


\author[1,7]{\fnm{Sanja} \sur{\v{S}\'{c}epanovi\'{c}}}\email{sanja.scepanovic@nokia-bell-labs.com}
\equalcont{These authors contributed equally to this work.}

\author[1,2]{\fnm{Sagar} \sur{Joglekar}}\email{sagarjoglekar@gmail.com}
\equalcont{These authors contributed equally to this work.}

\author[3]{\fnm{Stephen} \sur{Law}}\email{stephen.law@ucl.ac.uk}

\author[1,6,8]{\fnm{Daniele} \sur{Quercia}}\email{daniele.quercia@nokia-bell-labs.com}

\author[1,9]{\fnm{Ke} \sur{Zhou}}\email{ke.zhou@nokia-bell-labs.com}

\author[4,5]{\fnm{Alice} \sur{Battiston}}\email{ali.battiston@gmail.com}

\author[5]{\fnm{Rossano} \sur{Schifanella}}\email{rossano.schifanella@unito.it}

\affil[1]{\orgname{Nokia Bell Labs}, \orgaddress{\city{Cambridge}, \country{UK}}}

\affil[2]{\orgname{Intercom}, \orgaddress{\city{London}, \country{UK}}}

\affil[3]{\orgname{University College London}, \orgaddress{\city{London}, \country{UK}}}

\affil[4]{\orgname{Ofcom}, \orgaddress{\city{London}, \country{UK}}}

\affil[5]{\orgname{University of Turin}, \orgaddress{\city{Turin}, \country{Italy}}}

\affil[6]{\orgname{Kings College London}, \orgaddress{\city{London}, \country{UK}}}

\affil[7]{\orgname{University of Oxford}, \orgaddress{\city{Oxford}, \country{UK}}}

\affil[8]{\orgname{Politecnico di Torino}, \orgaddress{\city{Turin}, \country{Italy}}}

\affil[9]{\orgname{University of Nottingham}, \orgaddress{\city{Nottingham}, \country{UK}}}


\abstract{
Urban greenery is often linked to better health, yet findings from past research have been inconsistent. One reason is that official greenery metrics measure the amount or nearness of greenery but ignore how often people actually may potentially see or use it in daily life. To address this gap, we introduced a new classification that separates \emph{on-road greenery}, which people see while walking through streets, from \emph{off-road greenery}, which requires planned visits. We did so by  combining aerial imagery of Greater London and greenery data from OpenStreetMap with quantified greenery from over 100,000 Google Street View images and accessibility estimates based on 160,000 road segments. We linked these measures to 7.45 billion medical prescriptions issued by the National Health Service and processed through our methodology. These prescriptions cover five conditions: diabetes, hypertension, asthma, depression,  and anxiety, as well as opioid use.  As hypothesized, we found that green on-road was more strongly linked to better health than four widely used official measures. For example, hypertension prescriptions dropped by 3.68\% \sss{in wards with on-road greenery above the median citywide level compared to those below it.}
If all below-median wards reached the citywide median in on-road greenery, prescription costs could fall by up to \pounds 3.15 million each year. These results suggest that greenery seen in daily life may be more relevant than public yet secluded greenery, and that official metrics commonly used in the literature have important limitations.

}

\keywords{urban greenery, greenery exposure, urban nature, health, medical prescriptions, NDVI, WHO target, Natural England, ESA WHO, Google Street View, Open Street Map, road accessibility, public health}



\maketitle

Urban greenery is often assumed to support better health. Studies suggest links between exposure to greenery and lower rates of depression, asthma, and cardiovascular disease~\cite{MACKERRON2013992, bratman2015nature, Seresinhe2015, shepley2019impact, bratman2019nature}. Theories such as biophilia~\cite{biophiliahypothesis}, attention restoration~\cite{kaplan1995restorative}, and prospect-refuge~\cite{appleton1996experience} offer ways to explain these effects. 

Yet evidence remains mixed, partly because of how greenery is measured. Many studies use the Normalized Difference Vegetation Index (NDVI), which indiscriminately captures green surfaces from an aerial perspective~\cite{pettorelli2013normalized}, including greenery that may be private, distant, or not visible from the street. Three other commonly used measures based on green space targets from the World Health Organization (WHO and ESA WHO) or Natural England (NE) focus on coarse-grained proximity to parks or green areas~\cite{battiston2024need} without accounting for whether these spaces are part of people's daily experiences. As a result, these metrics may misrepresent how greenery may potentially impact everyday life.

To address this, we introduce a new classification that distinguishes between greenery seen along streets (\emph{on-road greenery}) and greenery found in parks or other public areas requiring a visit (\emph{off-road greenery}). This aims to reflect the difference between greenery people pass by during routine activities, and greenery they must go out of their way to access. We developed this classification by combining high-resolution aerial imagery (25 cm) of Greater London from the London Green Cover Map \cite{GreenCover2016} (commonly used for NDVI) to initially detect the presence of greenery (Datasets~\ref{methods_NDVI}). We then incorporated data on the location of green spaces from OpenStreetMap (OSM) \cite{OpenStreetMap} to delineate public parks and other designated green areas (Datasets~\ref{methods_OSM}). To assess \sss{on-road} visibility, we used over 100,000 Google Street View (GSV) images to identify greenery observable from street level (Datasets~\ref{methods_gsv}). Finally, we refined the classification to account for accessibility by mapping greenery to over 160,000 road segments, filtered using road centrality measures to ensure they reflect realistically traversable routes (Datasets~\ref{methods_osmeridian}).

For comparability, we computed four official greenery metrics: NDVI, WHO, ESA WHO, and NE. We derived NDVI using the London Green Cover Map \cite{GreenCover2016} (Methods~\ref{methods_calculating_NDVI}), and calculated the three proximity-based metrics of WHO and ESA WHO (Methods~\ref{methods_WHO_targets}), and NE (Methods~\ref{methods_natural_england}) based on the presence of green spaces identified in OSM and the ESA World Cover dataset~\cite{zanaga_daniele_2021_5571936}, following a previously validated methodology~\cite{battiston2024need}.

\begin{figure*}[t!]
    \centering
    \includegraphics[width=.65\textwidth]{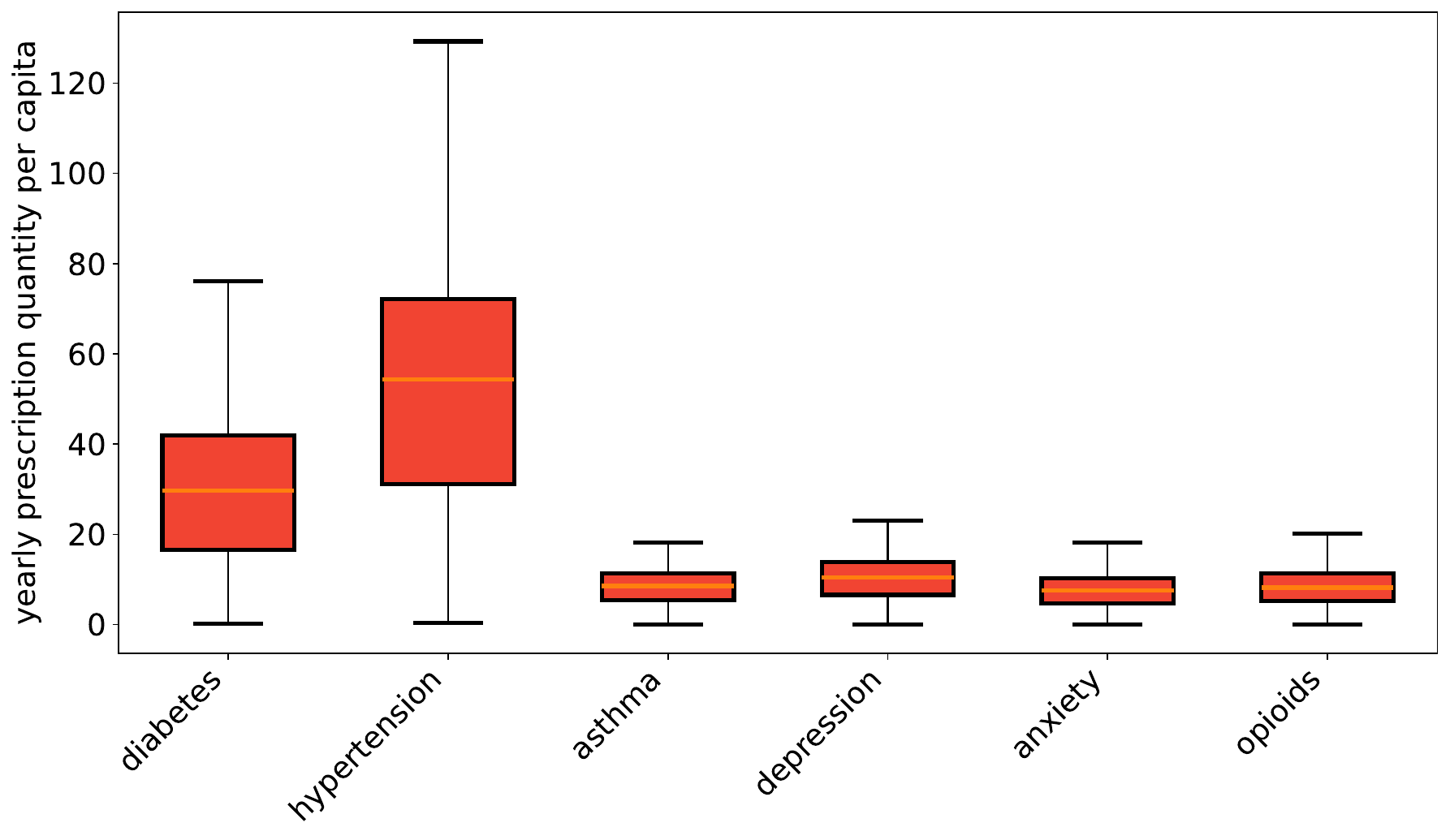}
    \caption{{\textbf{Distribution of yearly prescription quantity per capita for six health outcomes across London wards: diabetes, hypertension, asthma, depression, anxiety, and opioid prescriptions.} Metabolic conditions had the highest prescription rates, followed by antidepressants, while asthma had the lowest.}}
    \label{fig_dist_prescriptions}
\end{figure*}

We then linked these six greenery measures to per capita prescription rates for \sss{five} common and costly conditions influenced by environmental and social factors (Figure~\ref{fig_dist_prescriptions}): diabetes, depression, asthma, hypertension, anxiety~\cite{astell2014neighborhood,bodicoat2014association,sorensen2022air,li2022associations,leng2020exploring}, \sss{as well as} opioid use~\cite{schifanella2020spatial}. These rates were derived by assigning 7.45 billion prescriptions issued by the National Health Service (NHS)~\cite{PracticeLevelPrescribing,NHSBSA_EPD_Resource} 
to the conditions and mapping them from general practitioner practices to the wards they served, following the validated methodology outlined in \cite{scepanovic2024medsat} (Methods~\ref{methods_prescriptions}).

We tested the relationship between greenery and health using geographically weighted regression~\cite{jiang2018does,su2017area}, which models how associations vary across space, and propensity score matching~\cite{rosenbaum1983central}, which helps reduce bias when estimating causal effects from observational data.

\section{Results}

\begin{table}[t!]
    \caption{\textbf{Associations between greenery measures and prescription rates.} Associations between per capita prescription rates and the four official greenery measures (top), and our two proposed metrics (middle), along with the projected annual reduction if all wards in the city achieved at least 5.5\% on-road greenery. Average Treatment Effects (ATEs) are shown with standard errors.}
    \label{tab_all}
    \centering
    \footnotesize
    \begin{tabular}{llllllll}
    \toprule
     & diabetes & hypertens.\ & asthma & depression & anxiety & opioids & total \\
     \bottomrule
    \multicolumn{8}{l}{} \\
    \multicolumn{8}{l}{The four official metrics} \\
    \midrule
    
    NDVI&  
    \cellcolor{gray!10}\phantom{-}1.18{\tiny $\pm$.51} &
    \cellcolor{gray!10}\phantom{-}1.99{\tiny $\pm$.79} & 
    \cellcolor{gray!10}\phantom{-}0.86{\tiny $\pm$.43} & 
    \cellcolor{gray!10}\phantom{-}0.91{\tiny $\pm$.55} & 
    \cellcolor{gray!10}\phantom{-}0.19{\tiny $\pm$.25} & 
    \cellcolor{gray!10}-0.79{\tiny $\pm$.49} &
    \cellcolor{orange!10}\phantom{-}\textbf{1.90}{\tiny $\pm$.45} \\

   ESA WHO& 
   \cellcolor{gray!10}-0.49{\tiny $\pm$.33} & 
   \cellcolor{gray!10}-0.91{\tiny $\pm$.65} & 
   \cellcolor{gray!10}\phantom{-}0.83{\tiny $\pm$.40} & 
   \cellcolor{gray!10}-0.75{\tiny $\pm$.27} & 
   \cellcolor{gray!10}-1.99{\tiny $\pm$.85} & 
   \cellcolor{gray!10}-0.73{\tiny $\pm$.46} & 
   \cellcolor{gray!10}-0.93{\tiny $\pm$.56} \\
    
    WHO& 
    \cellcolor{orange!10}\phantom{-}\textbf{1.70}{\tiny $\pm$.48} & 
    \cellcolor{gray!10}\phantom{-}2.39{\tiny $\pm$.95} & 
    \cellcolor{gray!10}\phantom{-}2.68{\tiny $\pm$.98} & 
    \cellcolor{orange!10}\phantom{-}\textbf{3.08}{\tiny $\pm$.60} & 
    \cellcolor{orange!10}\phantom{-}\textbf{3.11}{\tiny $\pm$.46} & 
    \cellcolor{orange!10}\phantom{-}\textbf{3.23}{\tiny $\pm$.39} & 
    \cellcolor{gray!10}\phantom{-}1.40{\tiny $\pm$.63} \\
    
   NE& 
   \cellcolor{orange!10}\phantom{-}\textbf{2.45}{\tiny $\pm$.73} & 
   \cellcolor{orange!10}\phantom{-}\textbf{2.13}{\tiny $\pm$.69} & 
   \cellcolor{orange!10}\phantom{-}\textbf{2.29}{\tiny $\pm$.52} & 
   \cellcolor{orange!10}\phantom{-}\textbf{2.09}{\tiny $\pm$.75} & 
   \cellcolor{gray!10}\phantom{-}1.81{\tiny $\pm$.78} & 
   \cellcolor{gray!10}\phantom{-}1.49{\tiny $\pm$.64} & 
   \cellcolor{orange!10}\phantom{-}\textbf{1.78}{\tiny $\pm$.68} \\

     \bottomrule
    \multicolumn{8}{l}{} \\
    \multicolumn{8}{l}{Our two metrics} \\
    \midrule
   on-road & 
    \cellcolor{blue!10}\textbf{-3.01}{\tiny $\pm$.29} & 
    \cellcolor{blue!10}\textbf{-3.68}{\tiny $\pm$.33} & 
    \cellcolor{blue!10}\textbf{-2.32}{\tiny $\pm$.45} & 
    \cellcolor{blue!10}\textbf{-2.59}{\tiny $\pm$.51} & 
    \cellcolor{blue!10}\textbf{-2.92}{\tiny $\pm$.67} & 
    \cellcolor{blue!10}\textbf{-3.18}{\tiny $\pm$.39} & 
    \cellcolor{gray!10}-0.90{\tiny $\pm$0.48} \\

    off-road & 
    \cellcolor{gray!10}0.17{\tiny $\pm$.22} & 
    \cellcolor{gray!10}1.66{\tiny $\pm$.68} & 
    \cellcolor{gray!10}-1.56{\tiny $\pm$.72} & 
    \cellcolor{gray!10}-2.10{\tiny $\pm$.84} & 
    \cellcolor{gray!10}-0.84{\tiny $\pm$.44} & 
    \cellcolor{gray!10}-0.47{\tiny $\pm$.39} & 
    \cellcolor{gray!10}-1.26{\tiny $\pm$.53} \\
     
     \bottomrule
    \multicolumn{8}{l}{} \\
    \multicolumn{8}{l}{Reduction in prescriptions} \\
    \midrule
    quantity (M)\ & 
    -5.59 & -10.82 & -1.04 & -1.44 & -1.21 & -1.50 & -37.25 \\
    
    cost (£M) & 
    -1.14 & -0.80 & -0.37 & -0.12 & -0.15 & -0.26 & -3.15 \\
    
    \bottomrule
    \multicolumn{8}{l}{\footnotesize{Significant ATEs ($p<0.01$) are bolded.}}
    \end{tabular}
\end{table}

Our main result was that the four widely-used official measures showed inconsistent or even adverse associations (top panel of Table~\ref{tab_all}), and that only \emph{on-road greenery}, defined as vegetation visible along public streets, was consistently associated with lower prescription rates across all tested health conditions  (mid panel of Table~\ref{tab_all}). 

Next, we go into the details by spelling out three key aspects: how the four official measures perform, why they are insufficient, and how our two metrics offer a more nuanced perspective.


\mbox{ } \\
\textbf{How the four official measures perform.} We controlled for covariates identified in prior research on geo-spatial health\cite{browning2022greenspace,dendup2018environmental,brody1987trends,eisenman2019urban}, which include building density, deprivation index, median age, and ethnicity. \sss{This modeling is detailed in Methods \ref{methods_predicting}.} The top panel of Table~\ref{tab_all} shows the average treatment effects (ATEs) from linking each official greenery metric to prescriptions per capita. 
ATEs compare areas exposed to each greenery type \sss{(i.e., with above the citywide median greenery)} to matched control areas, using propensity score matching to reduce confounding bias. 
Higher values indicate more prescriptions; and significant values (in bold in Table~\ref{tab_all}) suggest a robust association. Our models showed that traditional greenery metrics produced inconsistent associations with health outcomes. NDVI and ESA WHO targets had mostly insignificant results. More surprisingly, the WHO and NE targets were positively associated with prescription rates for several conditions. For instance, the WHO target was linked with increased prescriptions for mental health conditions and opioids, and the NE target with metabolic, respiratory, and antidepressant prescriptions. 
\sss{Other studies also linked green space to higher prescription rates. For example, \citet{astell2022urban} found more antidepressant prescriptions with more open grass in Australia. They argued such areas were less compact, leading to more car travel, or could be hard to reach or unsafe, limiting time spent in nature. \citet{hyam2020greenness} saw more antidepressant and anxiolytic prescriptions with summer-greenness in Scotland. They suggested mown parkland, while green, might not offer enough social benefit or environmental perks, and argued that diverse, well-kept green spaces like street trees were more helpful. \citet{becker2022green} reported more opioid deaths with greater tree cover in the US, explaining this by historic natural resource jobs in heavily wooded regions like Appalachia. Lastly, \citet{aerts2020residential} noted more asthma prescriptions with grass cover. They linked this to allergic reactions from grass pollen and fungi, and less exposure to varied natural environments. See Section S1 in Supplementary Information for a full review of these conflicting findings.}



\begin{figure*}[t!]
    \centering
    \includegraphics[width=.48\textwidth]{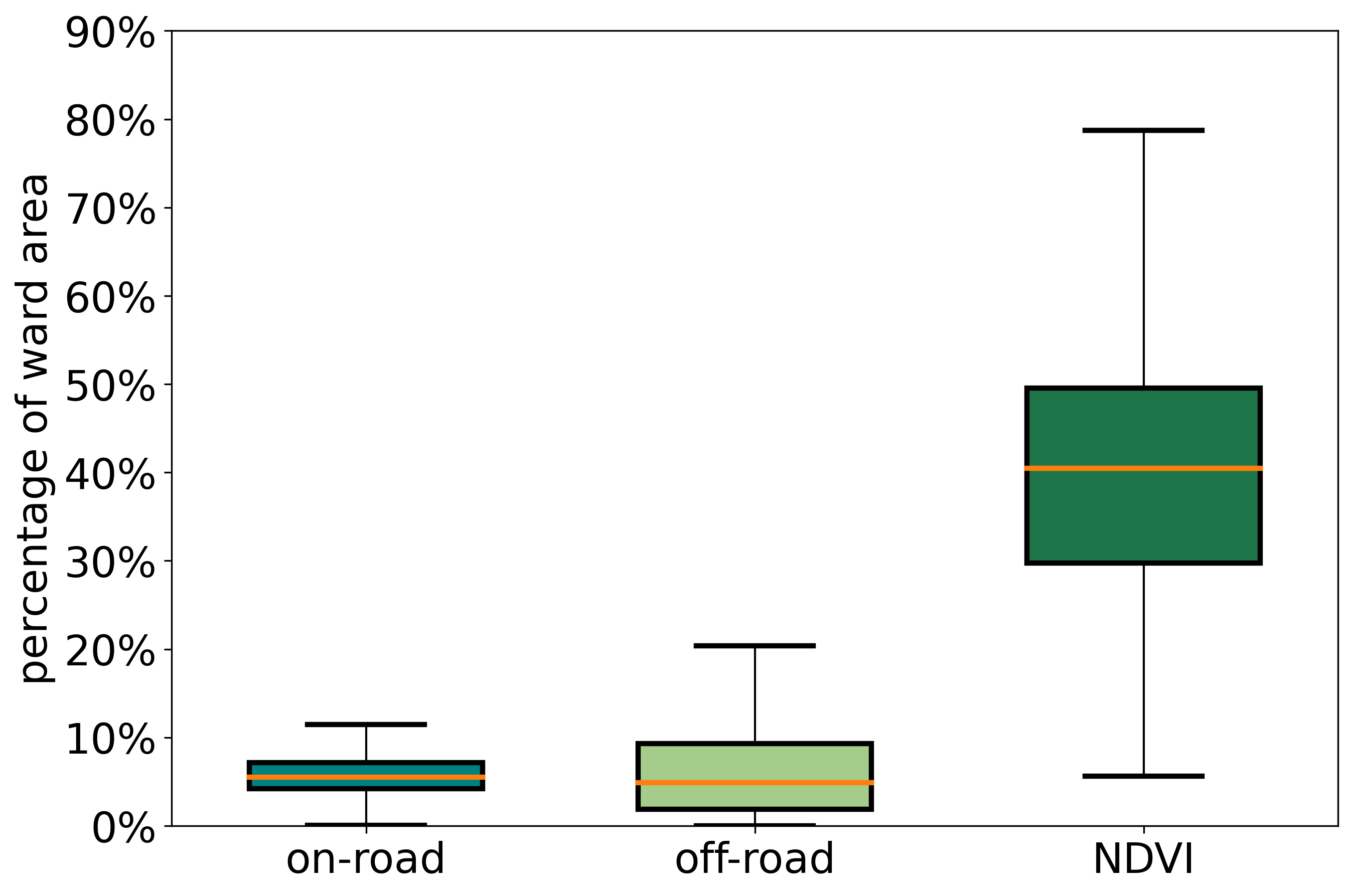}
    \includegraphics[width=.48\textwidth]{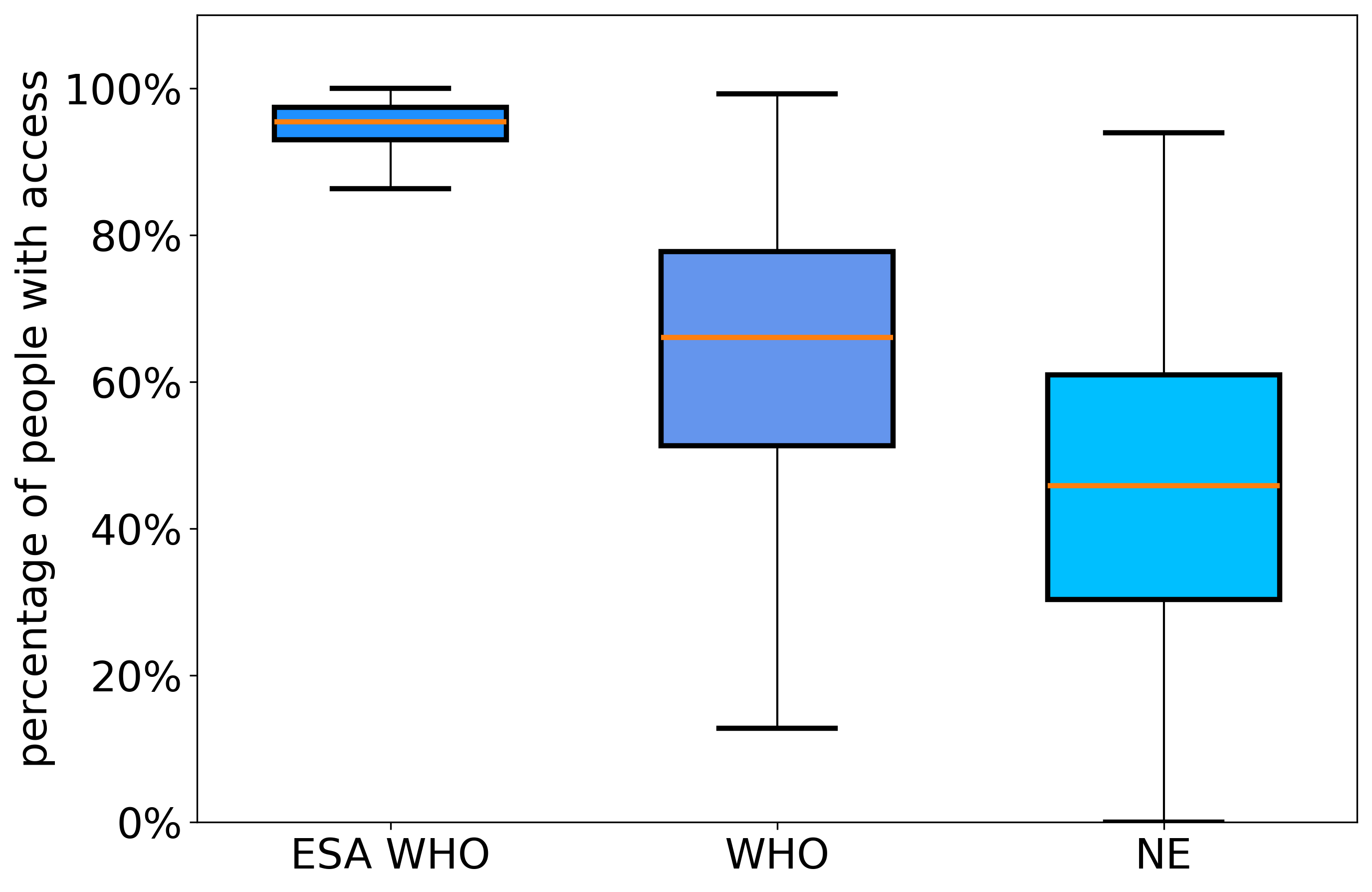}
    \caption{\textbf{Distribution of the six greenery measures across London wards.} The \emph{left} panel compares coverage from our two proposed metrics with the overly inclusive NDVI: on-road greenery averages 6\%, off-road greenery 8\%, while NDVI reaches as much as 41\%. The \emph{right} panel shows the proportion of residents meeting the three proximity-based targets: ESA WHO (95\%), WHO (64\%), and NE (45\%).}
    \label{fig_dist_greenery_metrics}
\end{figure*}

\mbox{ } \\
\textbf{Why the official measures are insufficient.} These inconsistencies may arise from two key limitations. First, both NDVI and ESA WHO conflate public and private greenery, which affects assessments of accessibility: private green areas such as farmland or gardens are not equally available to all residents. The distribution of greenery across wards, as measured by NDVI and ESA WHO (Figure~\ref{fig_dist_greenery_metrics}), unrealistically suggests that most areas in London meet the ESA WHO target, highlighting the overly inclusive nature of these metrics. NDVI, in particular, tends to be inflated in peripheral areas where large tracts of forest or agricultural land may exist, even though these spaces are not necessarily accessible to the public.
 
\begin{figure*}[t!]
    \centering
    \includegraphics[width=.96\textwidth]{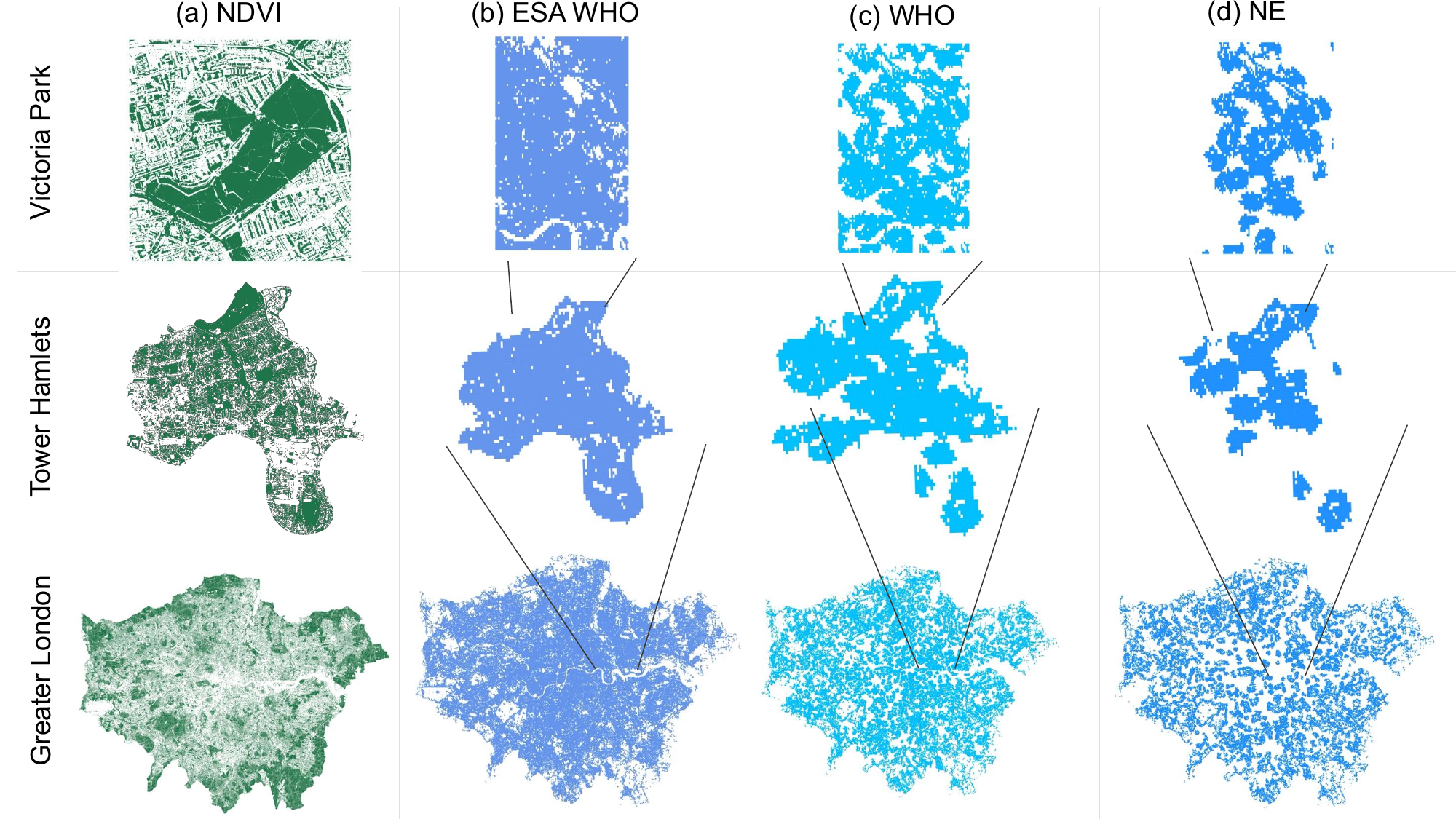}
    \caption{\textbf{Fine-grained spatial distributions of official greenery metrics.} These maps illustrate varying spatial scales: from Victoria Park to Tower Hamlets borough, and across Greater London.}
    \label{fig_official_greenery_maps}
\end{figure*}

Second, none of the four official metrics capture how easily people might encounter greenery during routine activities. The NE threshold considers distance but not visibility, and is therefore unmet in many central areas (Figures~\ref{fig_dist_greenery_metrics} and ~\ref{fig_official_greenery_maps}) despite the presence of street trees and other visible greenery. In contrast, the WHO and ESA WHO targets also consider distance but not actual travel routes, meaning they can be easily met by living near a large park, even if it is difficult to access. Lastly, NDVI does not account for street networks at all.

\mbox{ } \\
\textbf{How our two metrics offer a more nuanced perspective.} The mid panel in Table~\ref{tab_all} shows results for our two greenery types. We used the same causal inference method, controlling for the same covariates. Significant negative ATEs for on-road greenery indicate lower prescription rates across all health outcomes. Only on-road greenery had significant negative associations with prescription rates. The effects ranged from \(-3.68\%\) for hypertension to \(-2.32\%\) for asthma. Off-road greenery showed no significant associations. These findings support our hypothesis that both public access and routine visibility are critical. While both greenery types were public, only on-road greenery potentially aligned with  daily encounters. This practical exposure appears necessary for measurable health benefits, \sss{in agreement with~\citep{mears2021mapping,turunen2023cross}.}
\sss{Our results also align with studies showing eye-level greenery offers mental health benefits \citep{yu2024dynamic}, mitigates asthma risks~\citep{alcock2017land}, and road network-based greenery buffers correlate with lower diabetes rates \citep{mazumdar2021green} (see section S2 in Supplementary Information for detailed comparisons).}

To assess the practical impact of our findings, we estimated the potential reduction in prescriptions and associated costs if all wards currently below the median level of on-road greenery (5.51\%) were brought up to that median (bottom panel of Table~\ref{tab_all}). The model (detailed in Methods \ref{methods_prescriptions}) projects that such an increase could prevent approximately 37.25 million prescriptions annually, corresponding to cost savings of \pounds 3.15 million. These reductions span all studied health categories, with the largest effects observed for hypertension (10.82 million prescriptions) and the highest cost savings for diabetes (\pounds 1.14 million). Even other conditions
such as asthma and depression show gains, with cost savings from \pounds 120K to \pounds 290K.

\section{Discussion}

 While off-road greenery and widely used official greenery metrics were either unrelated or associated with higher prescription rates, on-road greenery stood out as the only measure showing consistent and beneficial effects. It consistently linked to lower prescription rates across all health conditions studied. This included significant associations with conditions such as hypertension, asthma, depression, and opioid use.

This main finding supports the idea that greenery must be both publicly accessible and encountered in daily routines to contribute to better health. The consistency and strength of the on-road effect suggest that visual and incidental contact with greenery in one's daily environment may also have a meaningful public health impact.

Traditional greenery metrics may fall short for two main reasons. First, they often conflate public and private greenery. For example, NDVI includes rooftops, gardens, and farmland, spaces not typically usable by the general public. Similarly, the ESA WHO target does not distinguish between accessible and inaccessible green areas. These metrics risk overestimating real-world exposure, especially in suburban or rural areas where private green space is abundant.

Second, they do not capture the visibility or frequency of exposure. NDVI, derived from satellite imagery, ignores street-level layout. WHO, ESA WHO, and NE targets rely on access within fixed distances but overlook whether people actually see or use those spaces. In contrast, our on-road greenery measure captures the greenery found along frequently used streets, making it more reflective of daily human experience.

The public health implications are substantial. If all areas below the citywide median for on-road greenery were brought up to that level, the resulting reduction in number of prescriptions could exceed 37 million annually. This would lead to savings of more than \pounds 3 million in prescription costs for the NHS, with the greatest reductions seen in treatments for hypertension and asthma.

\subsection*{Limitations}

Our study has two main limitations. First, although we used propensity score matching to reduce confounding and account for demographic and environmental variables, it remains an observational analysis. We cannot fully rule out the influence of unobserved factors that may affect both greenery levels and health outcomes. 

Second, while our proposed greenery measures improve upon existing ones, they still simplify complex realities. Human exposure to urban greenery depends on individual mobility, personal routines, and socio-spatial inequalities, which our metrics do not fully capture.

\subsection*{Future directions}

Despite these limitations, our findings provide actionable insights. Urban planners and public health professionals should consider not just how much green space is available, but whether it is integrated into the street network and potentially visible during daily activities. Street-level vegetation such as trees, hedges, and green facades may yield higher health returns than parks located far from residential or commercial routes.

Future work should continue refining exposure metrics to better capture lived experience, and explore how different communities interact with greenery based on age, occupation, mobility, and socio-economic status. Such research will help design greener cities that more equitably and effectively promote health.

\section{Datasets}\label{sec_methods_data}


Greater London consists of $625$ \emph{wards}, and $4,969$ \emph{Lower Layer Super Output Areas (LSOAs)}. Ward is the primary unit of English Electoral Geography and each ward has roughly the same number of voters. LSOAs are yet smaller geographical units created by the Office for National Statistics for statistical purposes and each LSOA has a population of approximately $1,500$ people. 

We conducted all our primary analyses at the level of wards. Although wards may not be completely homogeneous in terms of their characteristics, their size allows for the collection of a statistically significant number of data points (and GSV data were too sparse for the analysis at lower levels than that). To ensure robustness and minimze the effects of the Modifiable Area Unit Problem (MAUP) \cite{fotheringham1991modifiable} on our results, we then repeated all the analyses at the LSOA level (except for \emph{on-road greenery}, which we could calculate from NDVI at the LSOA level, but not from GSV). All LSOA-level analysis results are presented in Supplementary Information.


In order to study the relationship between greenery and urban outcomes quantitatively, we collected data from three main types of sources:
\begin{enumerate}
    \item \textbf{Greenery data}: to measure all types of \emph{greenery}, we used the NDVI London Green Cover Map derived from high resolution aerial imagery (Section~\ref{methods_NDVI}); to measure \emph{on-road greenery} in an alternative manner, we downloaded Google Street View images (Section~\ref{methods_gsv}); to asses \emph{on-road greenery accessibility}, we employed road centrality data from the OS Meridian Line (Section~\ref{methods_osmeridian}); and finally, to separate \emph{accessible} from \emph{inaccessible  greenery}, we used Open Street Map parks data (Section~\ref{methods_OSM});
    \item \textbf{Medical prescriptions data}: to estimate prescription prevalence per capita for six individual conditions (diabetes, hypertension, depression, anxiety, opioids, asthma), as well as total, we used Open Prescriptions Dataset released freely by the NHS (Section \ref{methods_prescriptions}).
    \item \textbf{Control data}: for socioeconomic indicators, we used UK Census data (Section \ref{methods_census}).
\end{enumerate}

\subsection{NDVI Data: London Green Cover Map}
\label{methods_NDVI}
\begin{figure*}[hbt!]
    \centering
    \includegraphics[width=.777\linewidth]{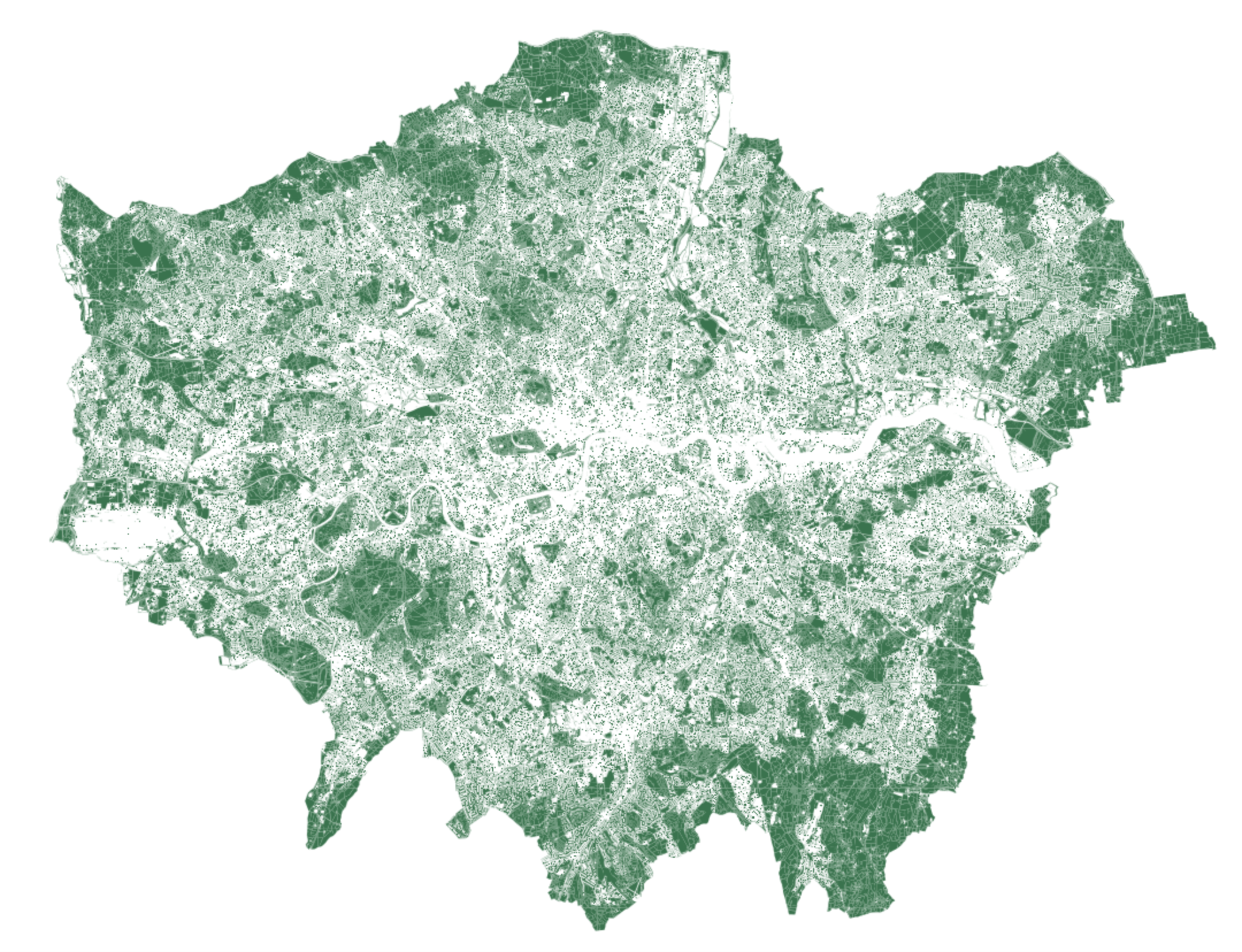}
    \caption{\textbf{Spatial distribution of the NDVI}. Source: London Green Cover.\label{figNDVI}}
\end{figure*}

The amount of green cover in an area is commonly measured using remote sensing data, either from satellite or aerial images. Examples of satellite sources include Terra MODIS, Landsat series, Pléiades, KOMPSAT, and Planet. The resolution of the images varies from low (e.g., MODIS with 250–1000m) to medium (e.g., Landsat with 15 to 60m) to high (e.g., Planet with 50cm). With aerial images, i.e., those taken from a plane, even a higher resolution is possible (e.g., 10cm). While previous green cover maps for London were mostly derived from satellite images, for this study, we used the London Green Cover map, which can be accessed at this URL: \url{https://apps.london.gov.uk/green-cover}, and it is produced from a September 2016 aerial source of 25cm resolution \cite{greenlondond} of near-infrared aerial imagery (NDVI) combined with two  topography layers (Ordnance Survey MasterMap, and Verisk GeoInformation Group UKMap) to capture small areas such as gardens, green roofs, small parks, and tree canopy. The London Green Cover map is provided as a high-resolution vector layer, consisting of 467 tiles each covering approximately 2km x 2km (Figure \ref{figNDVI}). The map's accuracy is evaluated at 96.5\% \cite{greenlondond}.

\subsection{Open Street Map (OSM) parks data}
\label{methods_OSM}
We queried all areas in Greater London tagged as parks (\texttt{leisure=park}) and gardens (\texttt{leisure=garden}) from OpenStreetMap (OSM) via the Overpass API \cite{OSM_overpass}. The distinction between parks and gardens according to the OSM tagging guidelines is that a urban park is an \emph{``open space for recreational use, usually designed and in semi-natural state with grassy areas, trees and bushes''}, whereas gardens are \emph{``[\ldots] for the display, cultivation, and enjoyment of plants and other forms of nature.''}
At the time of the query in March 2023, there were $3,452$ parks and $14,289$ gardens of varying sizes, and these areas make up most  greenery within a city with the exception of e.g., cemeteries.

Naturally, the access to these green areas is sometimes restricted temporally within opening hours, or only available for certain groups in case of private and residential gardens. Such restrictions are also indicated using the \texttt{access} tag, which can indicate that the area is generally accessible to the public (\texttt{access=[yes | permissive]}), or not (\texttt{access=[no | private]}).
Of the 17741 parks and gardens 8487 (47.8\%) are tagged with information regarding their access. 

After a careful manual inspection, we found that all the parks and all the gardens that did not have the tag \texttt{access=[no | private]} were generally accessible for the public. Hence, we collated shapefiles for X gardens and Y parks in total, which we used for calculation of \emph{accessible  greenery}.

\subsection{Google Street View (GSV) Imagery}
\label{methods_gsv}
We collated front-facing street images in London using the Google Street View API accessed through \url{https://developers.google.com/maps/documentation/streetview#street-view-features-and-apis}. The images are facing the street instead of facing buildings. We followed the method described in the papers \cite{Law2018,law2019take} to construct a graph from the roads of London. We used the centroid of each road and its azimuth as parameters to collect this image dataset. In total we obtained 104,778 street images in London.

\subsection{OS Meridian London Street Network}
\label{methods_osmeridian}
The Ordnance Survey Meridian Line (OS Meridian) is an openly available simplified road network for the UK, which can be accessed at this URL: \url{https://data.london.gov.uk/dataset/ordnance-survey-opendata}. The dataset contains a total of $161,645$ street segments for London.

\subsubsection{WHO and ESA WHO Targets Data}
\label{methods_WHO_targets_data}
The World Health Organization's green accessibility guidelines recommend urban residents have access to at least 0.5–1 hectare of green space within 300 meters of their home. This study examined two interpretations of this: (1) access to a public green area of at least 0.5 hectares within a 5-minute walk (\textit{WHO target}), and (2) exposure to a total of 0.5 hectares of green features within the same range (\textit{ESA WHO}). 
To assess these targets across London, we employed a grid-based approach \cite{battiston2024need} with a refined 3 arc-second resolution grid ($\sim$60 meters). Population data were sourced from the Global Human Settlement Layer (version R2023A) \cite{ghs_pop}, and walking distances were calculated using OSRM's walking profile \cite{luxen-vetter-2011}. Data on public green spaces came from OpenStreetMap (May 2022) \cite{OpenStreetMap} for the \textit{WHO target} and from the 2020 ESA World Cover dataset \cite{zanaga_daniele_2021_5571936} for \textit{ESA WHO}, following green feature classification in \cite{battiston2024need}. Aggregation to LSOA or ward level used weighted averages based on population, with adjustments for partial grid-cell overlaps with geographic boundaries.

\subsection{Natural England (NE) Target Data}
\label{methods_natural_england_data}
In 2010, Natural England introduced a target recommending access to a public green area of at least 2 hectares within a 5-minute walk of residential locations \cite{england2010nature}. This study assessed compliance with this target using the grid-based method in \cite{battiston2024need}, enhanced by a finer 3 arc-second resolution grid ($\sim$60 meters) derived from the Global Human Settlement Layer (version R2023A) \cite{ghs_pop}. Walking distances were calculated using OSRM’s walking profile \cite{luxen-vetter-2011}, and public green area data were sourced from OpenStreetMap (OSM). 
Green accessibility was evaluated at the grid level, and results were aggregated to LSOA or ward levels using population-weighted averages. For grid cells partially overlapping geographic units, population weights were adjusted proportionally based on the intersection area.
\subsection{Open Prescriptions Data}
\label{methods_prescription_data}
\begin{figure*}[t!]
    \centering
    \includegraphics[width=.7\textwidth]{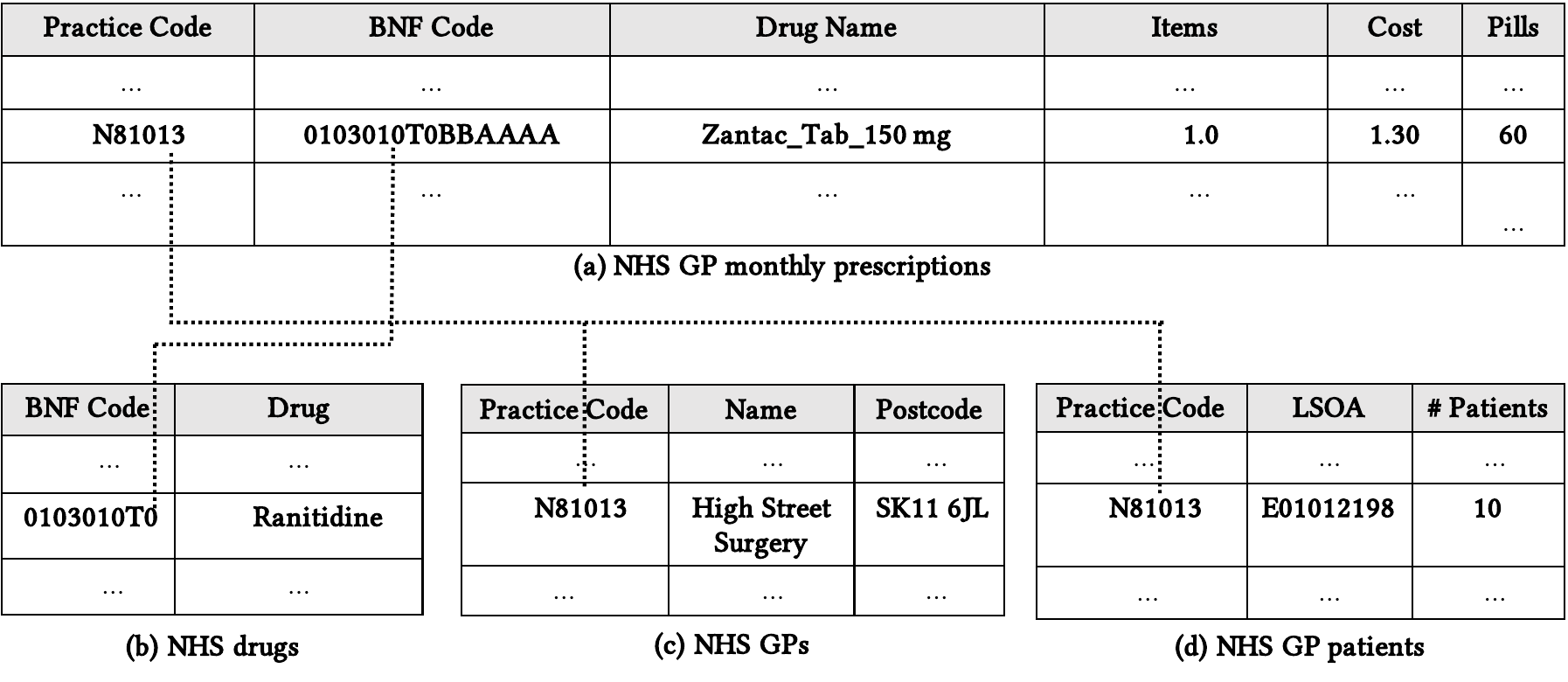}
    \caption{
        The four NHS Datasets: \emph{(a)} GP monthly prescriptions; \emph{(b)} Drugs; \emph{(c)} GPs; and \emph{(d)} Patients. Each monthly prescription in dataset \emph{(a)} was translated into a drug name based on the BNF code for which dataset \emph{(b)} offered the corresponding drug (preparation name). The prescription was also geographically mapped using the GP code for which dataset \emph{(c)} provided the location.  To then map the prescription at the level of census LSOAs, we computed the fraction of the GP's patients who lived in each LSOA from dataset  \emph{(d)}.}
    \label{Fig:data_link}
\end{figure*}

\begin{table*}[t!]
    \centering
    \footnotesize 
    \resizebox{0.8\linewidth}{!}{\begin{tabular}{|l|l|l|l|}
            \hline
            \textbf{BNF Code} &\textbf{Drug name} & \textbf{BNF Code} &\textbf{Drug name} \\
            \hline
            0601021V0 & Tolazamide & 0601023V0 & Metformin Hydrochloride/Rosiglitazone \\
            0601011A0 & Insulin Aspart & 0601023Z0 & Metformin Hydrochloride/Vildagliptin \\
            0601023A0 & Acarbose & 0601021P0 & Glipizide \\
            0601023X0 & Sitagliptin & 0601021E0 & Chlorpropamide \\
            0601012V0 & Insulin Glargine & 0601021H0 & Glibenclamide \\
            0601023M0 & Miglitol & 0601023AN & Empagliflozin \\
            0601011L0 & Insulin Lispro & 0601023AM & Canagliflozin \\
            0601023S0 & Rosiglitazone & 0601023W0 & Metformin Hydrochloride/Pioglitazone \\
            0605020E0 & Desmopressin Acetate & 0601023U0 & Nateglinide \\
            0601023B0 & Pioglitazone Hydrochloride & 0601023Y0 & Exenatide \\
            0601023AH & Saxagliptin/Metformin & 0607010B0 & Bromocriptine \\
            0601012X0 & Insulin Detemir & 0601023AI & Lixisenatide \\
            0601022B0 & Metformin Hydrochloride & 0212000AD & Colesevelam Hydrochloride \\
            0601012AB & Insulin Glargine/Lixisenatide & 0601021A0 & Glimepiride \\
            0601021M0 & Gliclazide & 0202080M0 & Hydrochlorothiazide/Potassium \\
            0601012Z0 & Insulin Degludec & 0601023AD & Metformin Hydrochloride/Sitagliptin \\
            0601023AX & Ertugliflozin & 0601012S0 & Isophane Insulin \\
            0601023AW & Semaglutide & 0601023AR & Empagliflozin/Metformin \\
            0601023AS & Albiglutide & 0601023AP & Canagliflozin/Metformin \\
            0601023AQ & Dulaglutide & 0601011R0 & Insulin Human \\
            0601023AV & Saxagliptin/Dapagliflozin & 0601023AL & Dapagliflozin/Metformin \\
            0601023R0 & Repaglinide & 0601023AJ & Alogliptin/Metformin \\
            0601023AK & Alogliptin & 0202010L0 & Hydrochlorothiazide \\
            0601021X0 & Tolbutamide & 0601023AF & Linagliptin/Metformin \\
            0601023AG & Dapagliflozin & 0601011P0 & Insulin Glulisine \\
            0601023AE & Linagliptin & 0601023AB & Liraglutide \\
            0601023AC & Saxagliptin & 0601023AA & Vildagliptin \\
            \hline          
    \end{tabular}}
    \caption{List of drug names associated with diabetes.}
    \label{tab:diabetesdrugs}
\end{table*}

The prescription data, which has been available since July 2010, is released monthly by NHS~\cite{PracticeLevelPrescribing,NHSBSA_EPD_Resource} and includes four files (Figure~\ref{Fig:data_link}):

\begin{enumerate}
    \item \emph{GP monthly prescriptions} (Figure~\ref{Fig:data_link}\emph{(a)}) -- These are anonymized prescriptions during a month in the whole England. Each prescription contains the drug's name, its British National Formulary (BNF) code~\cite{BNFglossary}, the practice code, the total number of items, total cost, and each item's quantity.

    \item \emph{Drugs}  (Figure~\ref{Fig:data_link}\emph{(b)}) -- Each row contains a drug name and its unique BNF code~\cite{BNFglossary}. We used DrugBank~\cite{knox2010drugbank} to identify which drug names are prescribed for diabetes and we used a list of pain drugs (opioids) curated in previous works~\cite{schifanella2020spatial,curtis2019opioid}.

    \item \emph{GPs} (Figure~\ref{Fig:data_link}\emph{(c)}) -- Each row contains a practice code, name, and full address. We filtered out closed or prison-hosted GPs, leaving us with 6924 GPs across England.

    \item \emph{Patients} (Figure~\ref{Fig:data_link}\emph{(d)}) -- Each row contains a practice code, census LSOA area code, and the number of the practice's patients in that area. We used this data to calculate the total number of primary care patients who live in a certain area: 

    \begin{equation}
        n_{pat}{(a)} = \sum_{gp} n_{pat}(gp,a)
        \label{formula:n-patients@a}
    \end{equation}
    where $n_{pat}{(gp,a)}$ is the number of $GP$'s patients who live in $a$. We found that the number of patients in a ward correlates strongly ($r=.92$) with the number of residents, supporting our choice of ward as a unit of spatial analysis. For each $gp$ and a given area $a$, we also  computed the corresponding \emph{fraction} of the $gp$'s patients:
    \begin{equation}
        \begin{aligned}
            f(gp,a) = \frac{ n_{pat}{(gp,a)}  } { n_{pat}{(gp)} }
        \end{aligned}
        \label{eq:GP_patients}
    \end{equation}

\end{enumerate}

\subsection{Socioeconomic Controls Data}
\label{methods_census}
The Office of National Statistics (ONS) conducts a nationwide census in the UK every 10 years, to measure socioeconomic, health, and cultural dynamics of the population. This data is available to the general public at spatial scales, including LSOAs. 

We collected data on \emph{median income, population density}, and \emph{average resident age} from the UK census, as our control variables. This variables are useful to understand the relationship between urban outcomes in London and the presence of greenery because they can mediate such relationships by relating with one or both of the factors. Income can be an indicator of the resources available for maintaining greenery. Also, several studies have linked economic deprivation of areas to their social and health conditions \cite{drukker2003children,zierler2000economic,ng2010effect,bursik1993economic,sun2011cross}. Population density can affect how greenery is distributed and perceived in a neighborhood. The link between population density (a good proxy for urbanization) and crime is peculiar: crime scales super-linearly with density until a particular point;  then it continues to  grow sub-linearly~\cite{chang2019larger}. Yet, in both growth regimes, density goes hand-in-hand with crime, and that is why we controlled for it as well. A lot of health conditions are associated with an aging population. Particularly, the prevalence of diabetes and chronic pain has a strong association with age \cite{kirkman2012diabetes,rustoen2005age}. We therefore controlled for the average age of the population in an area.

\section{Methods}
\label{methods_methods}


\subsection{Calculating NDVI}
\label{methods_calculating_NDVI}
To measure the total greenery ($g_{total\_NDVI}(a)$) for our study area $a$ (i.e., a ward or LSOA), we applied the standard approach of overlaying the NDVI London Green Cover map (\emph{NDVI}) with respective administrative boundaries and calculating the percent of pixels $p$ in green-areas to the total study area pixels:
\begin{equation}\label{eq_totalNDVI}
        g_{\textrm{total\_NDVI}}(a) = \frac{\sum_p ({p \in NDVI)}} {\sum_p ({p \in a})}.
\end{equation}

\subsection{Calculating WHO and ESA WHO targets}
\label{methods_WHO_targets}
The latest guidelines from the World Health Organization in terms of green accessibility in urban areas recommend that each urban resident should have access to at least 0.5-1 hectare of green space within 300 meters of their home. In this study, we examined two alternative interpretations of this recommendation: one proposing that each resident should have access to a public green area of at least 0.5 hectares within a 5-minute walk from their home, and the other suggesting that residents should be exposed to a total of 0.5 hectares of green features within a 5-minute walk of their home. Throughout our manuscript, we denote the former interpretation as the \textit{WHO target} and the latter as the \textit{ESA WHO}.
To calculate the proportion of the population meeting each target for every LSOA or Ward within the city boundary of London, we used the grid-based approach outlined in \cite{battiston2024need}. Unlike the original version of this approach, we used here a finer grid with a 3 arc-seconds resolution (approximately 60 meters).
Specifically, grid and population data were sourced from version R2023A of the Global Human Settlement population layer \cite{ghs_pop}, while walking distances between any two cells within the grid were computed using the walking profile of the routing engine OSRM \cite{luxen-vetter-2011}. Data on public green spaces were extracted from OpenStreetMap (OSM) (May 2022 extract) \cite{OpenStreetMap} for the \textit{WHO target} and from the 2020 ESA World Cover dataset \cite{zanaga_daniele_2021_5571936} for the \textit{ESA WHO target}. In both cases, the selection of green features followed the approach outlined in \cite{battiston2024need}. 
By integrating these datasets, we could evaluate the level of green accessibility for every cell in the grid and determine whether each cell meets the specified targets. Results at the grid level were then aggregated to the highest geographical level (either LSOA or Ward) by computing a weighted average, with weights corresponding to the resident population of each cell. When aggregating the data at the Wards/LSOAs level, for cells at the boundary of the geographic unit that did not fall completely in the unit itself, we only apportioned to the LSOA/Ward a number of people proportional to the size of the geographical intersection between the cell and the LSOA/Ward.

\subsection{Calculating Natural England (NE) target}
\label{methods_natural_england}
In 2010, the UK governmental body Natural England established a multi-level target system to monitor access to nature across various levels. In this study, we considered the first target of this system, which recommend immediate access to a public green area of at least 2 hectares (here operationalized as access within a 5-minute walk from a residential location) \cite{england2010nature}.
As with the \textit{WHO} and \textit{ESA WHO targets} (see \ref{methods_WHO_targets}), the computation of this metric followed the approach outlined in \cite{battiston2024need}, enhanced by the uses of the finer base grid with a resolution of 3 arc-seconds provided by version R2023A of the Global Human Settlement population layer \cite{ghs_pop}. Walking distances between any two cells within the grid were computed using the walking profile of the routing engine OSRM \cite{luxen-vetter-2011}. Data on public green areas were sourced from OpenStreetMap (OSM). By integrating these datasets, we could evaluate the level of green accessibility for every cell in the grid and determine whether each cell meets the specified targets. Aggregation from the grid level to the highest geographical level (either LSOA or Ward) was performed via a weighted average, with the population of each cell serving as the weighting factor. When aggregating the data at the Wards/LSOAs level, for cells at the boundary of the geographic unit that did not fall completely in the unit itself, we only apportioned to the LSOA/Ward a number of people proportional to the size of the geographical intersection between the cell and the LSOA/Ward.

\subsection{Calculating Prescription Prevalence Scores}
\label{methods_prescriptions}
We studied medical prescriptions for following six individual conditions: diabetes, hypertension, depression, anxiety, and asthma, as well as opioids.

To calculate prescriptions for a condition, we either used a list of curated drugs (such as in the case of diabetes; see Table \ref{tab:diabetesdrugs}, and opioids, for which we used the drug lists curated in~\cite{schifanella2020spatial,curtis2019opioid}), or we crawled the DrugBank to create such a list automatically. Drugbank\footnote{https://go.drugbank.com/} is an online database that curates information about active pharmacological ingredients and their associated conditions. At the time of the crawl, the website contained information about 9105 drug names. Each name is  associated with: one or more conditions (i.e., symptoms and diseases); one or more drug categories; and one Anatomical Therapeutic Chemical (ATC) code, which uniquely identifies the drug according to a classification system maintained by the World Health Organization (WHO). We linked each drug name found on DrugBank to one or multiple: 1) drug categories; 2) symptoms (the drug is supposed to treat); and 3) diseases.  For example, we linked  the drug name Ibuprofen\footnote{\url{https://go.drugbank.com/drugs/DB01050} } to:  1)  the categories anti-inflammatory preparations, non-steroids for topical use, and analgesics; 2) the symptoms headache, migraine, acute pain; and 3) a variety of diseases, from severe pain to fever to osteoarthritis. If a drug name was not associated with either of the three, then it was discarded from our list: from the initial  9105 drug names, we were left with 3013 names. Finally, we curated a list of all the drugs that are associated with a given condition $c$, by filtering for associated drugs from this remaining list of drugs. These drugs were then matched to their respective BNF codes from the prescriptions data set, for the purpose of counting the prescriptions related to the selected condition $c$.


For each area $a$, we estimated the number of prescriptions for the condition $c$ as:
\begin{equation}
N_c{(a)} = \sum_{GP \in a} N_c{(GP)} \cdot f{(GP,a)},
\label{formula:n-items@a}
\end{equation}
where $N_c{(GP)}$ is the total number of $GP$'s prescriptions for drugs in the condition's curated list, and $f{(GP,a)}$ is the fraction of $GP$'s patients who live in area $a$, inferred using Formula \ref{eq:GP_patients}. 
To calculate total prescriptions for area $a$, in Formula \ref{formula:n-items@a}, we considered all the medications prescribed by a GP for any condition.

To make the prescription metric comparable across areas with different population densities, we computed \emph{prescriptions quantity per capita}, as is typically done in medical studies~\cite{curtis2018openprescribing,curtis2019opioid},:
\begin{equation}
\tilde{N_c}{(a)} =\frac{ N_c{(a)} }{ N_c^{pat}{(a)} }, \
\label{eq:condition-items@a}
\end{equation}
where $N_c^{pat}{(a)} $ corresponds to the total number of patients that live in area $a$, as obtained using Formula \ref{formula:n-patients@a}.

\begin{figure*}[t!]
    \centering
    \includegraphics[width=.44\textwidth]{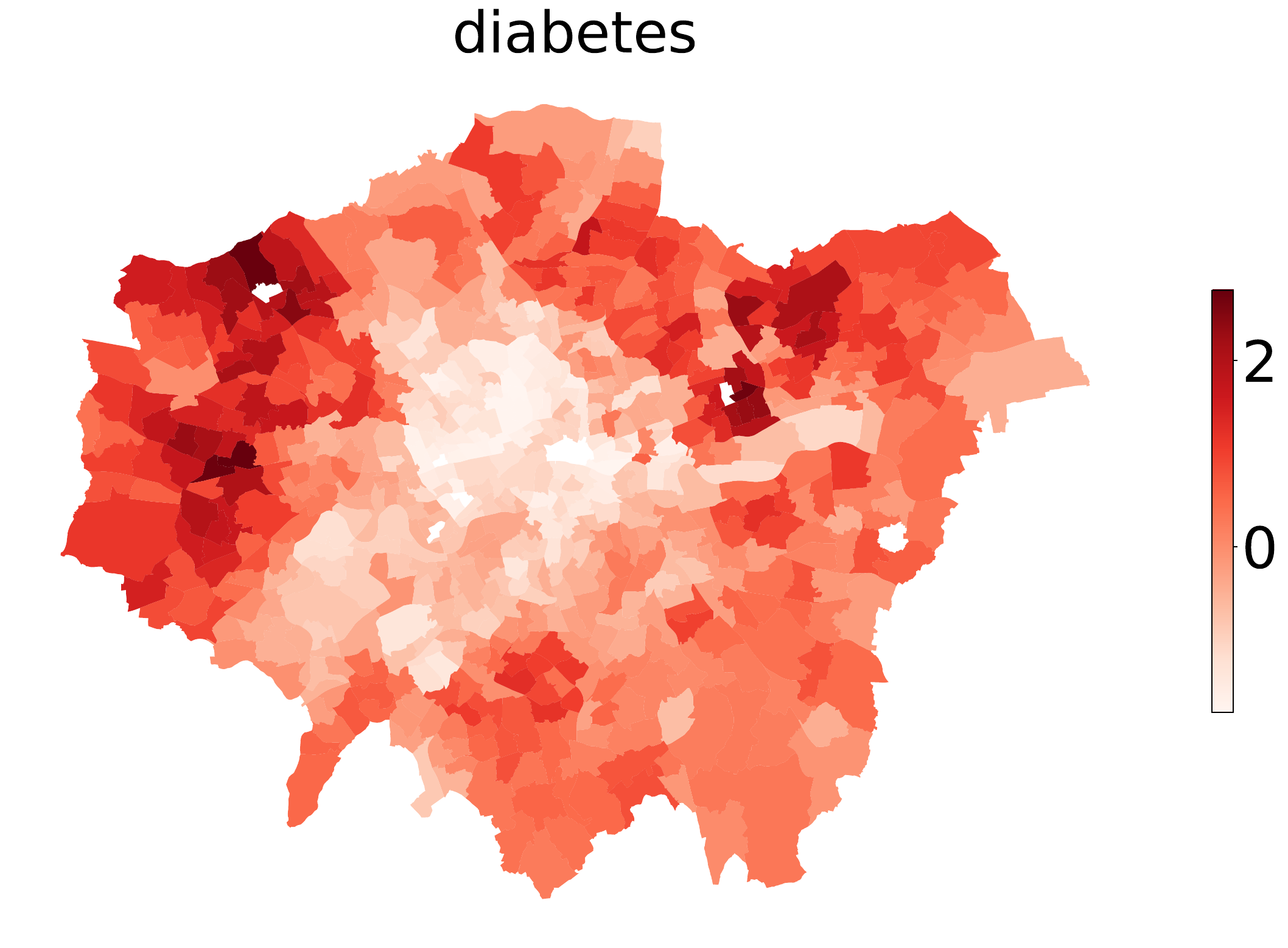}
    \includegraphics[width=.44\textwidth]{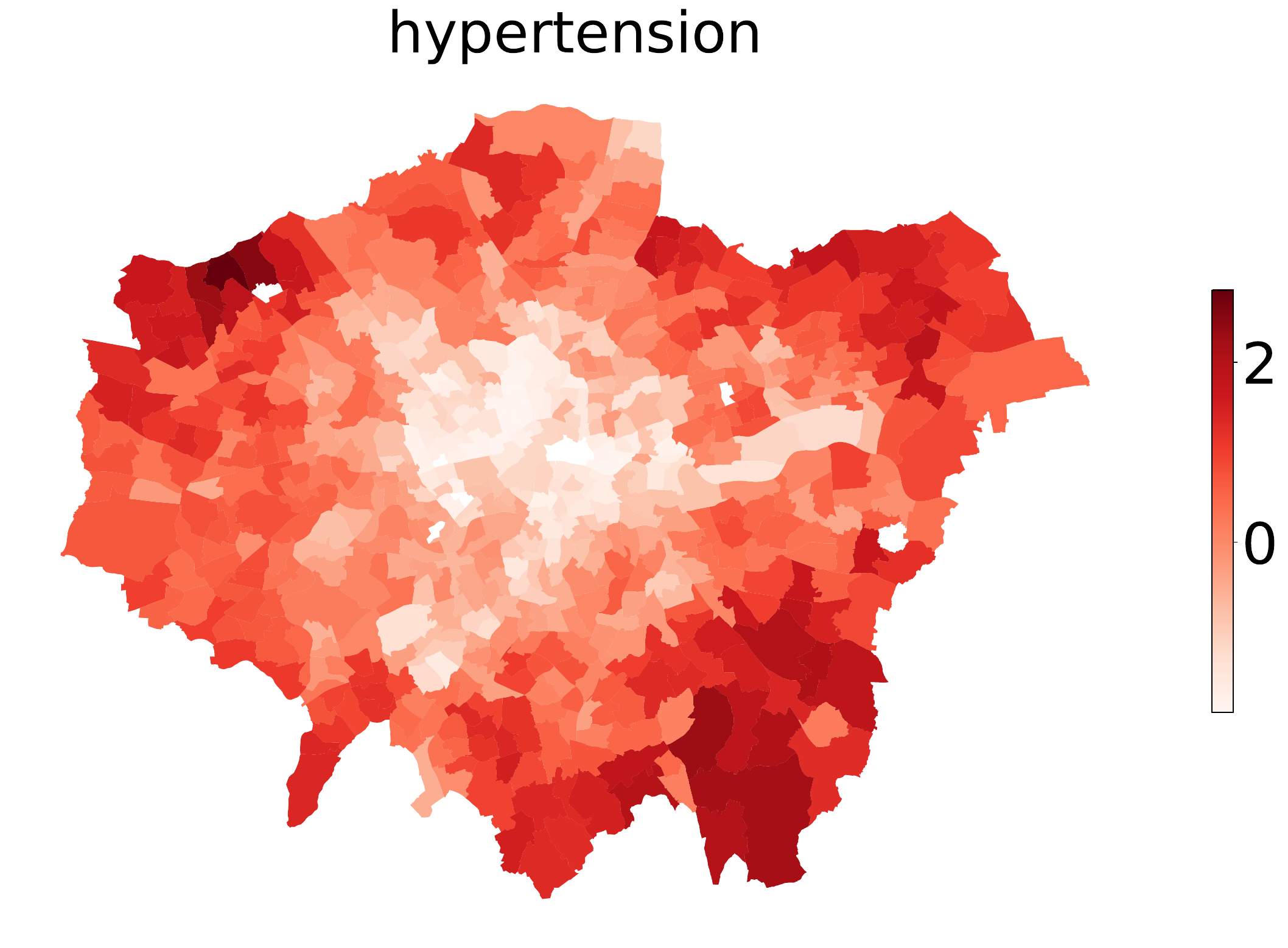}
    \includegraphics[width=.44\textwidth]{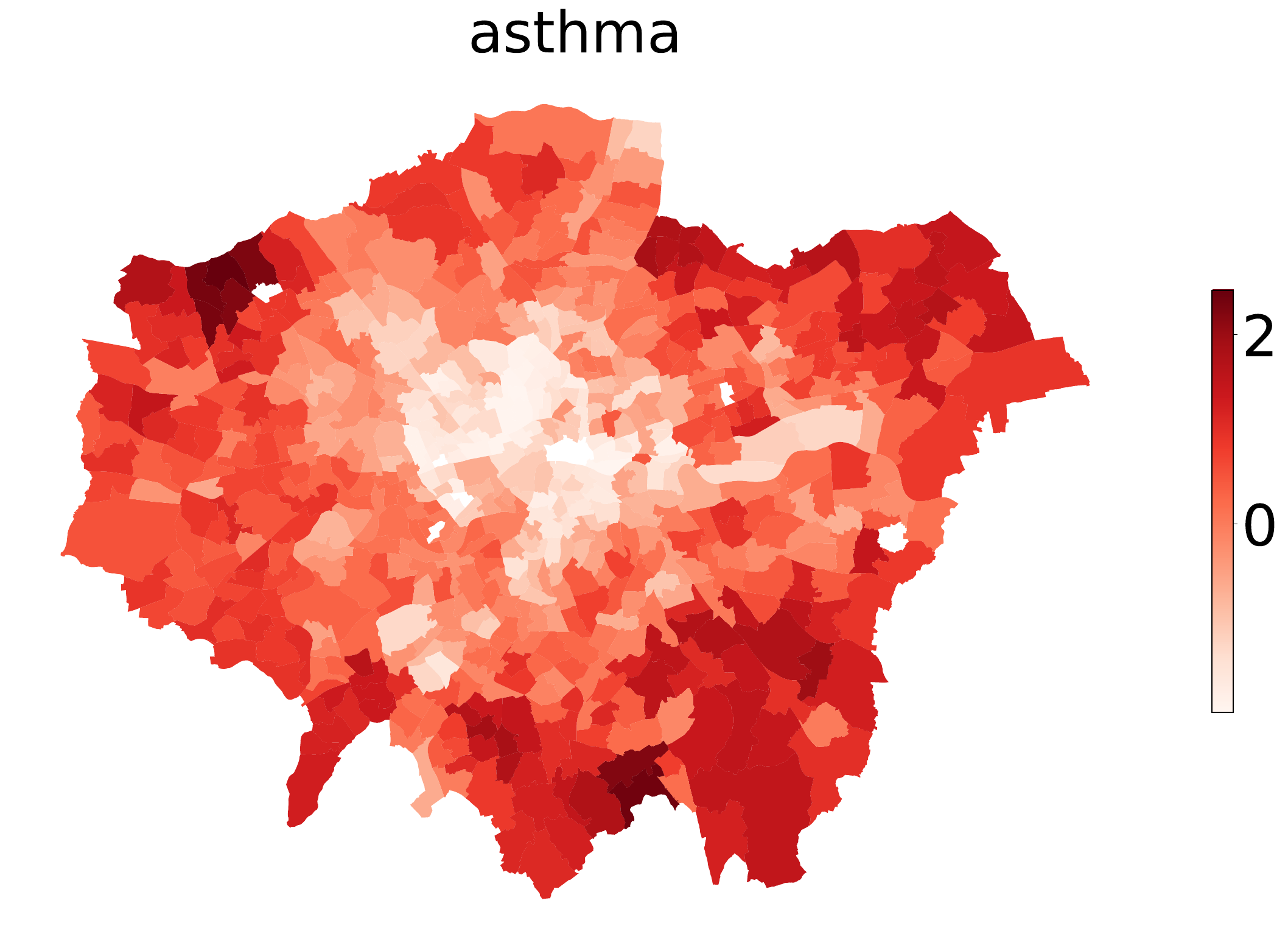}
    \includegraphics[width=.44\textwidth]{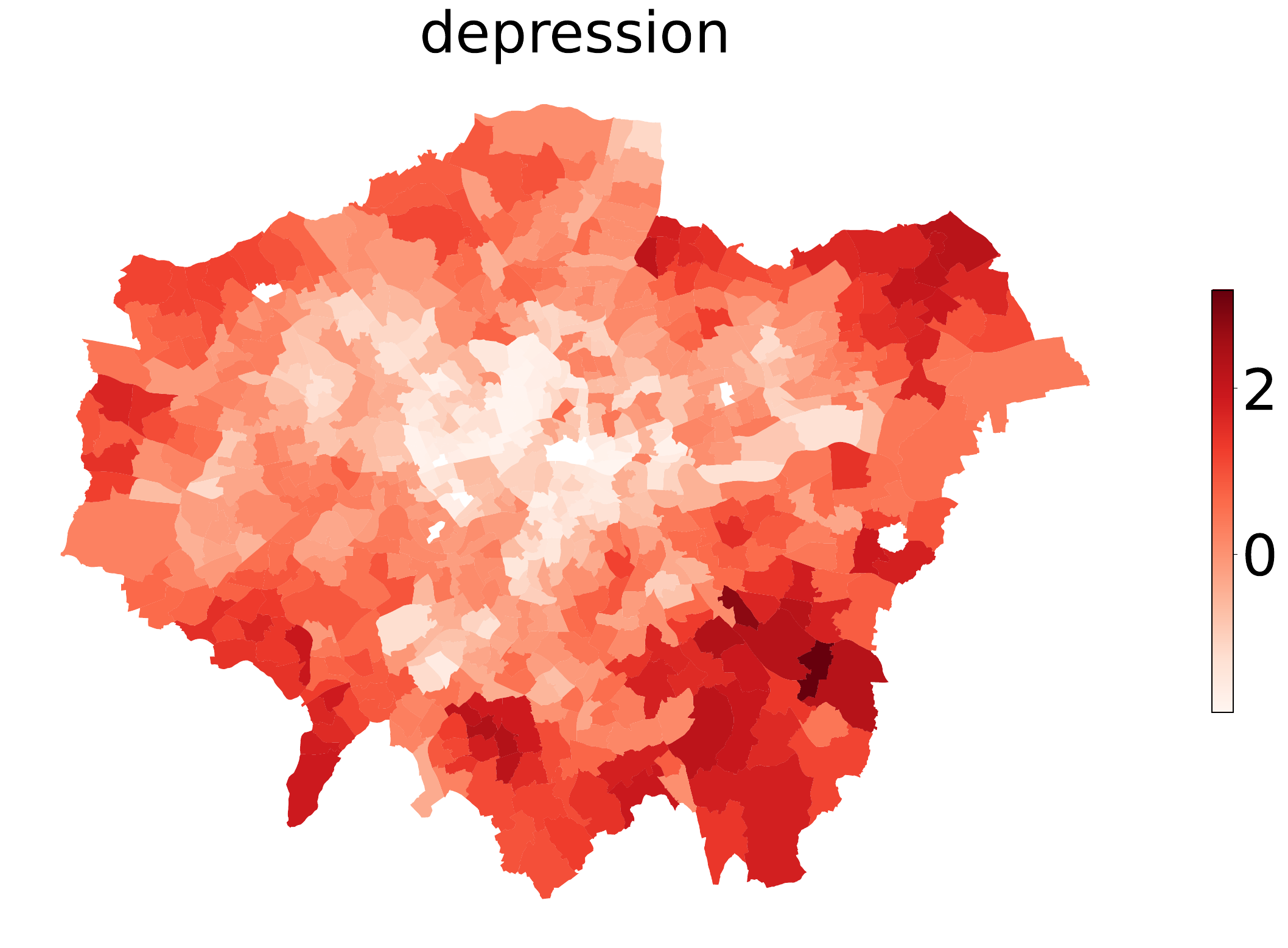}
    \includegraphics[width=.44\textwidth]{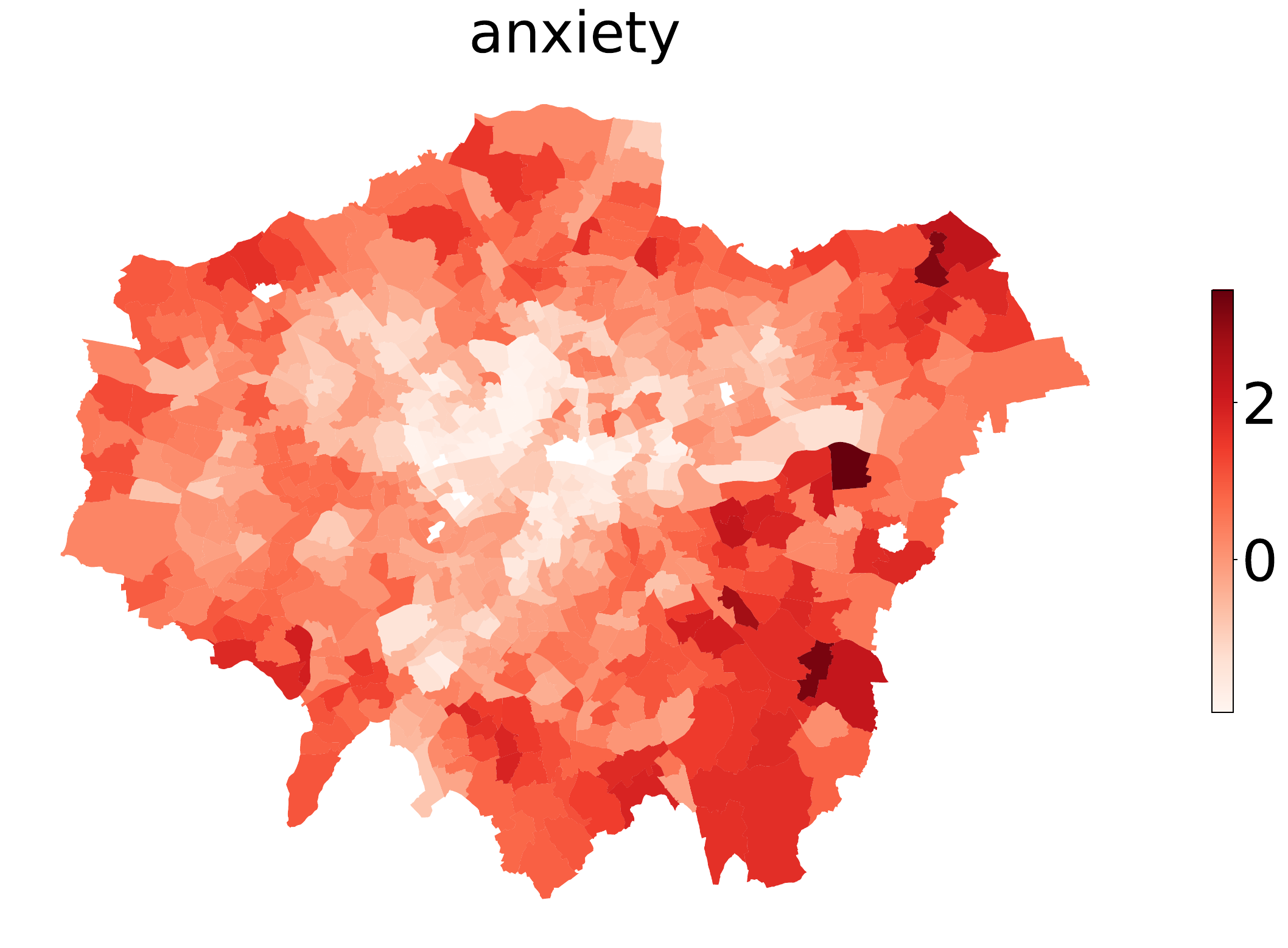}
    \includegraphics[width=.44\textwidth]{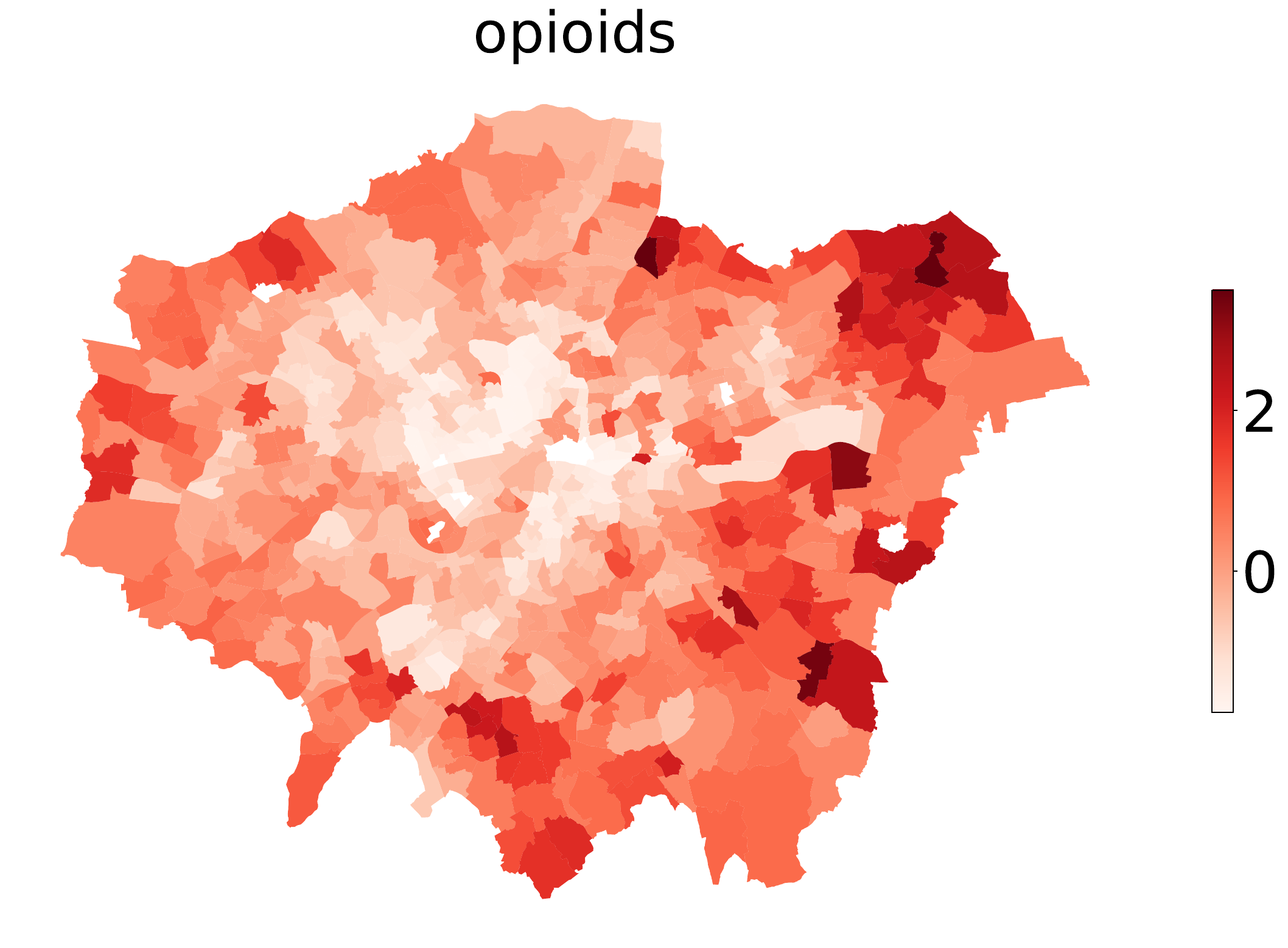}
    \includegraphics[width=.44\textwidth]{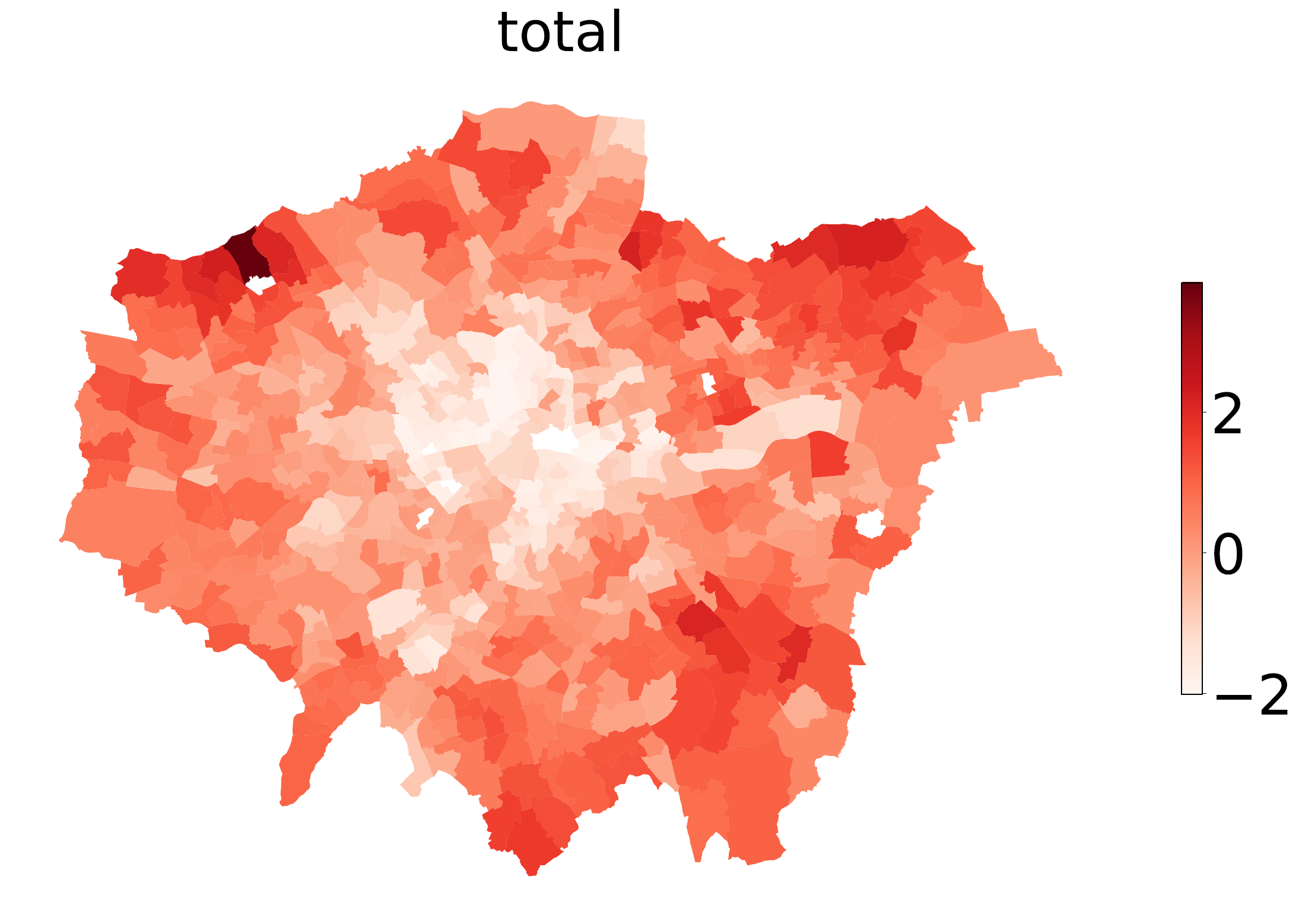}

    \caption{\textbf{Spatial distributions of prescription quantities per capita for diabetes, hypertension, asthma, depression, anxiety, and opioids and total, across London wards.} Standardized scores are shown, darker colors indicate higher prescription quantities.}
    \label{fig_spatial_dist_prescriptions}
\end{figure*}

The spatial distributions of yearly quantities per capita ($\tilde{N_c}{(a)}$) across London wards are shown in Figure~\ref{fig_spatial_dist_prescriptions}.

\paragraph{{Calculating reduction in prescriptions and costs}}

The potential reductions in prescriptions or costs (\(R\)) can be estimated using the PSM method's average treatment effects (ATEs). These represent the average difference in prescriptions per capita between wards with greenery levels above the city's median and those below. The calculation is as follows:

\[
R = \sum_{a \in \text{ControlGroup}} \tilde{N_c}{(a)} \cdot \text{ATE},
\]

where \(R\) is the total estimated reduction (in prescriptions or costs), $\tilde{N_c}{(a)}$ are total prescriptions or costs in area \(a\) from the control group, and \(\text{ATE}\) are the average treatment effect of greenery.

\subsection{Calculating Off-road Greenery}
\label{methods_off_road}
To calculate public greenery, we intersected all open parks and gardens with \emph{NDVI greenery}. 
From this public greenery, we then extracted the green spaces beyond the 10m buffers around the street segments, naming it \emph{off-road greenery}.
This enabled us to identify green areas that are publicly accessible, well-maintained, and often visited intentionally rather than encountered incidentally like on-road greenery.

\subsection{Calculating On-road Greenery}
\label{methods_on_road}
With \emph{on-road greenery} ($g_{\textrm{on-road}}$), we aimed at quantifying the greenery that citizens are routinely exposed to e.g., whilst they commute or go about their daily activities. 

In the first instance, we extracted the \emph{on-road greenery} from public greenery introduced at the previous step.
GSV is suggested to offer a better measure for the visible greenery compared to NDVI. Hence, we also tested using GSV as an alternative data source for \emph{on-road greenery}. Then we could compare the two type of measures for \emph{on-road greenery}: the one based on NDVI ($g_{\textrm{on-road}^{NDVI}}$), and the other one based on GSV ($g_{\textrm{on-road}^{GSV}}$).

For both, we first considered the road network (Section~\ref{methods_osmeridian}), and extracted information about greenery present around the roads as follows. For each road $i$, we created a 10-meter buffer $B_i$ on either side of the road segment. The 10m buffer accounts for the average width of a footpath in London (2m) and the distance until the verge of a road. Most roadside trees managed by the planning authority would be found inside our buffer.\footnote{\url{https://assets.publishing.service.gov.uk/government/uploads/system/uploads/attachment_data/file/341513/pdfmanforstreets.pdf}}


For $g_{\textrm{on-road}^{NDVI}}$, we the selected all the NDVI pixels inside a buffer, and computed the fraction of those pixels to the total study area $a$ (in which $i$ is located) pixels:
\begin{equation}\label{eq_on_roadNDVI}
        g_{\textrm{on-road}^{NDVI}}(i) = {\frac{\sum_p ({p \in NDVI \land p \in B_i)} } {\sum_p ({p \in a)}}}.
\end{equation}
The rest of the pixels that were found outside the street buffer, we termed \emph{ greenery}, and used to calculate the other two complementary forms of greenery later.
For $g_{\textrm{on-road}^{GSV}}$, we took all the Street View images inside the buffers $B_i$ and computed the fraction of pixels classified to be green in each image:
\begin{equation}\label{eq_on_roadGSV}
        g_{\textrm{on-road}^{GSV}}(i) = {\sum_{\textrm{GSV\_image} \in B_i} \frac{\sum_p ({p \in \textrm{GSV\_image} \land p \textrm{ is green}}) } {\sum_p ({p \in \textrm{GSV\_image})}}}.
\end{equation}

The pixel classification was done with a pre-trained SegNet model \cite{badrinarayanan2016segnet}, a deep encoder-decoder architecture for multi-class pixel-wise segmentation that was previously used to assess the quality of Street Views \cite{ye2019daily}, aesthetics of urban scenes \cite{joglekar2020facelift}, and associate daily access to greenery with economic performance \cite{ye2019measuring}.

For both types of \emph{on-road greenery}, we also developed alternative versions of the measure accounting for the road daily accessibility. The idea is that the more central a road is, the more visited it is likely to be \cite{hillier2007space,penn1998configurational,hillier2005network}, and the more its greenery is likely to be routinely experienced by citizens. To account for road accessibility, we used the space syntax choice measure (we describe how it is calculated in following section) and integrated it in the above metrics as follows. For each road segment $i$, first, we computed the local choice value $c_i$, log-transformed it due to the skewed distribution, 
and normalized it on a scale of 0 to 100. To remove any inaccessible road segments such as dead ends or cul-de-sacs we floored $c_i$ to $0$ if $c_i \leq 1$. We then took the weighted average of the $g_{\textrm{on-road}}(i)$ score ($g_{\textrm{on-road}^{GSV}}(i)$ or $g_{\textrm{on-road}^{GSV}}(i)$) by giving more importance to central roads:
\begin{equation}\label{eq_area_on_road}
        g_{\textrm{on-road}}(a) = \frac{\sum_{i \in a} log(c(i)) \cdot g_{\textrm{on-road}}(i) }{\sum_{i \in a} log(c(i))}.
    \end{equation}

\paragraph{Calculating Road Centrality}
We hypothesized that the benefits of greenery depend heavily on its centrality. To measure centrality of \emph{on-road greenery}, we employed an approach developed in prior research \cite{ye2019measuring}. This approach estimates accessibility and potential pedestrian flow for a given street or road segment by incorporating Space Syntax measures of accessibility. Space Syntax comprises theories and techniques that analyze the spatial configuration and pedestrian movement of urban areas in linking space and society \cite{hillier1989social, hillier2007space, penn1998configurational}.

A commonly used measure in Space Syntax is ``space syntax choice'' or angular weighted betweenness, which identifies street segments with the highest through-movement potential, and is calculated as:
\begin{equation}
c(i) = \sum_{j=1}^n \sum_{k=1}^n \frac{g_{jk}(i)}{ g_{jk}},
\end{equation}
where $i$ is the street segment under consideration, $j$ and $k$ are street segments other than $i$, $n$ is the total number of roads in the area, $g_{jk}(i)$ is the number of paths between $j$ and $k$ that pass through $i$, and $g_{jk}$ is the number of total paths between $j$ and $k$.
This weighted measure has been found to correlate highly with aggregate pedestrian movement \cite{hillier2005network}.

We calculated space syntax choice using the UCL \texttt{depthmapX} tool \cite{varoudis2013space, gil2015space} with a radius of 500m, as our main interest was pedestrian-scale access to greenery \cite{ye2019measuring}.

\begin{figure*}[t!]
    \centering
    \includegraphics[width=.44\textwidth]{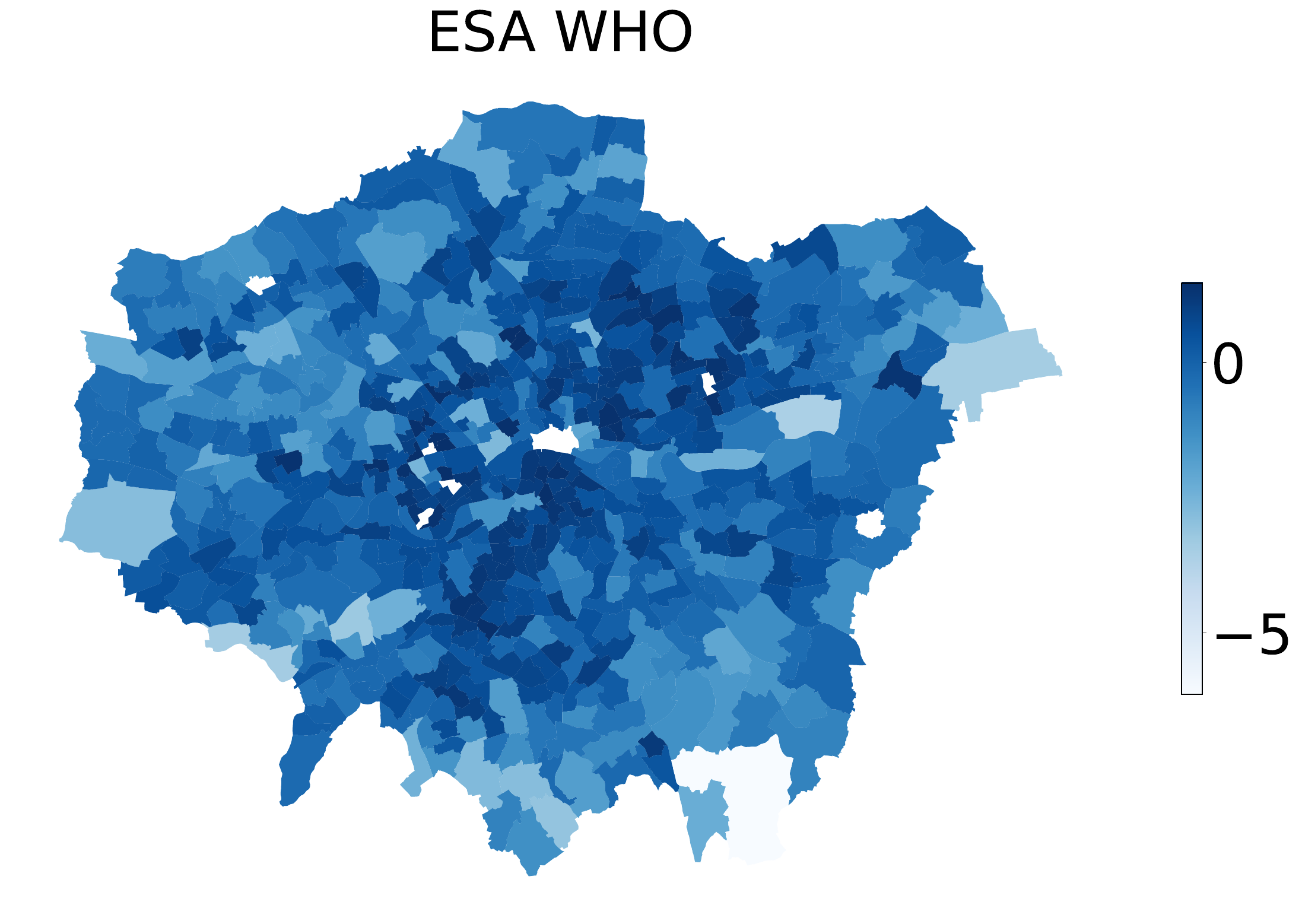}
    \includegraphics[width=.44\textwidth]{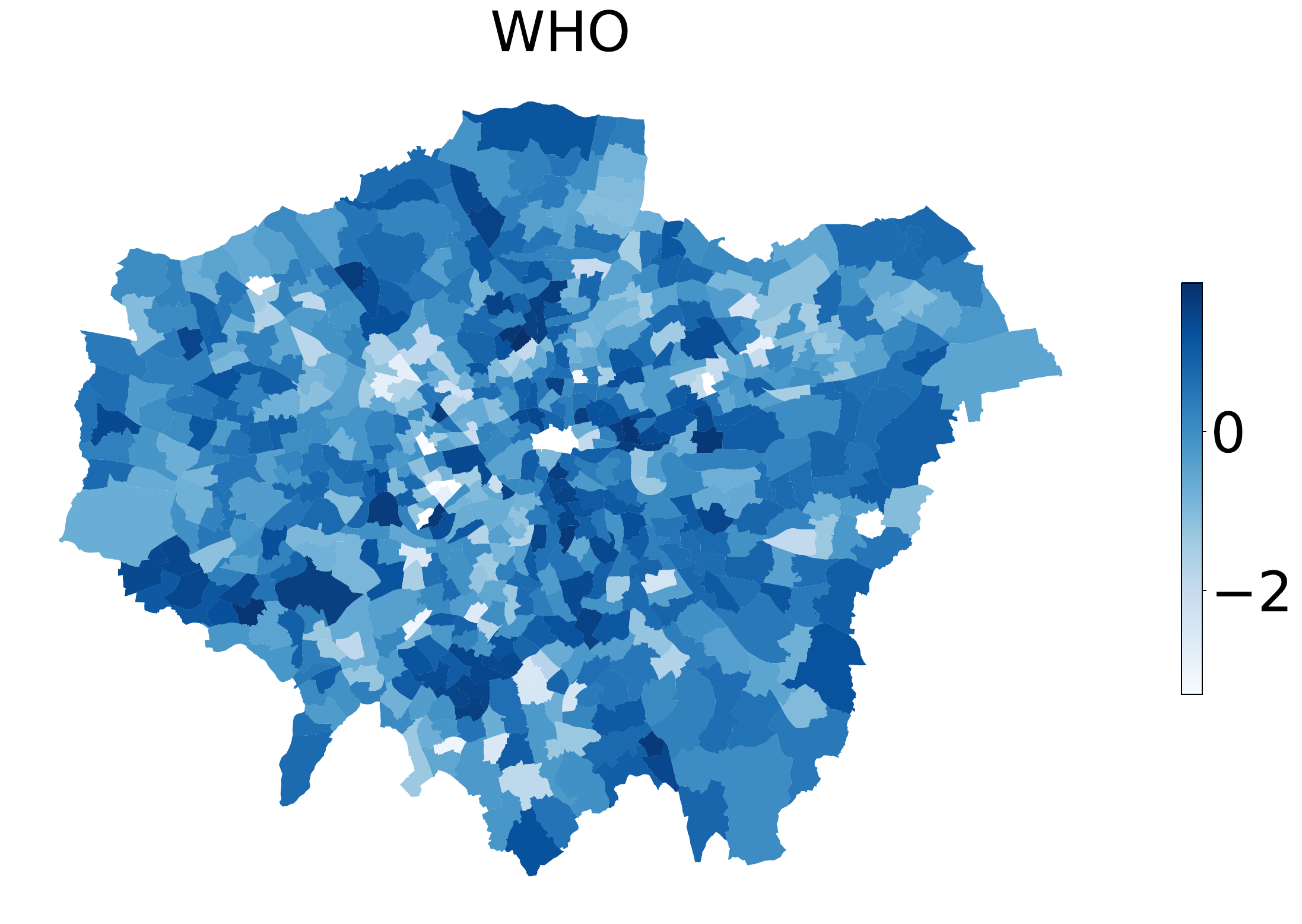}
    \includegraphics[width=.44\textwidth]{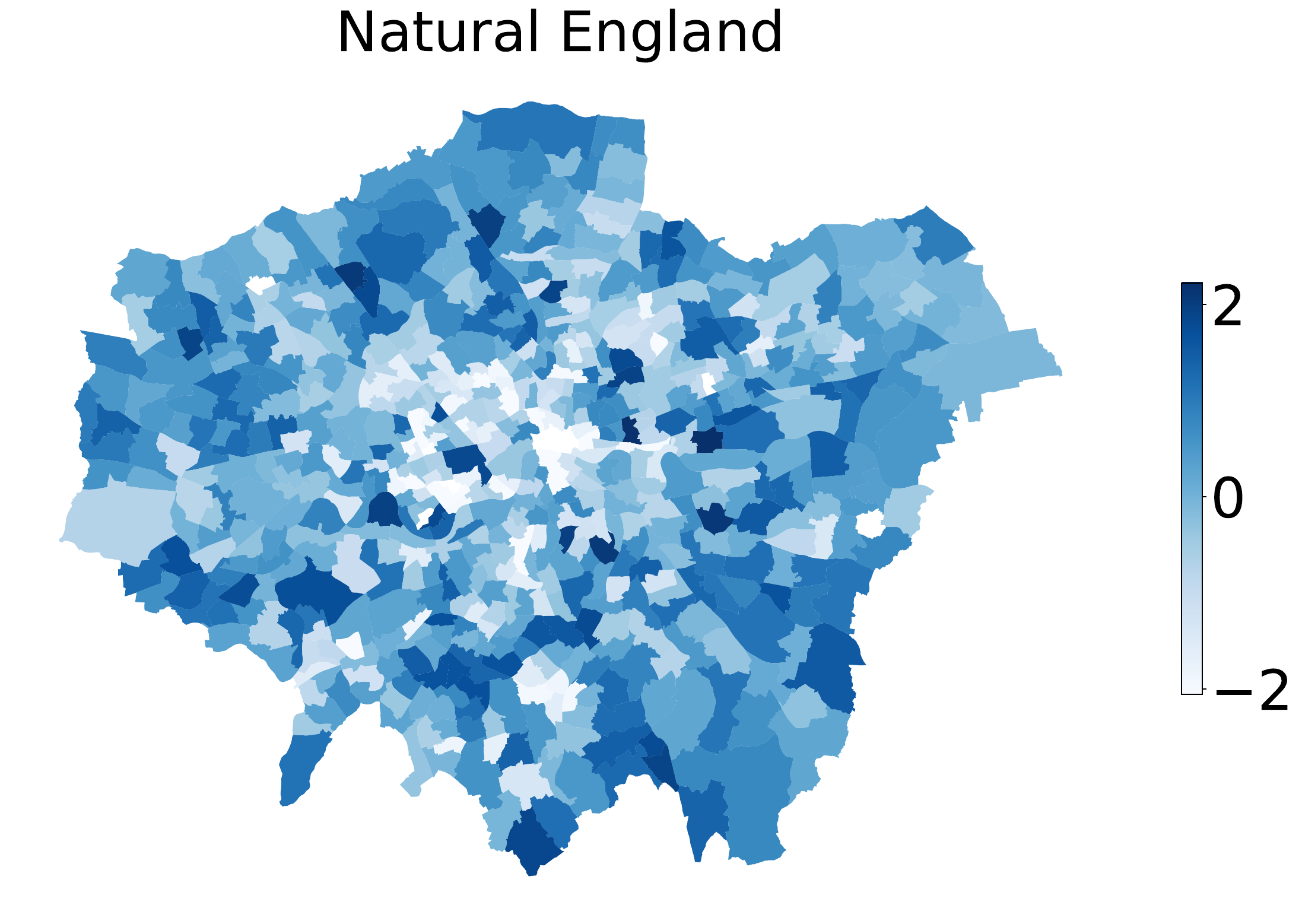}
    \includegraphics[width=.44\textwidth]{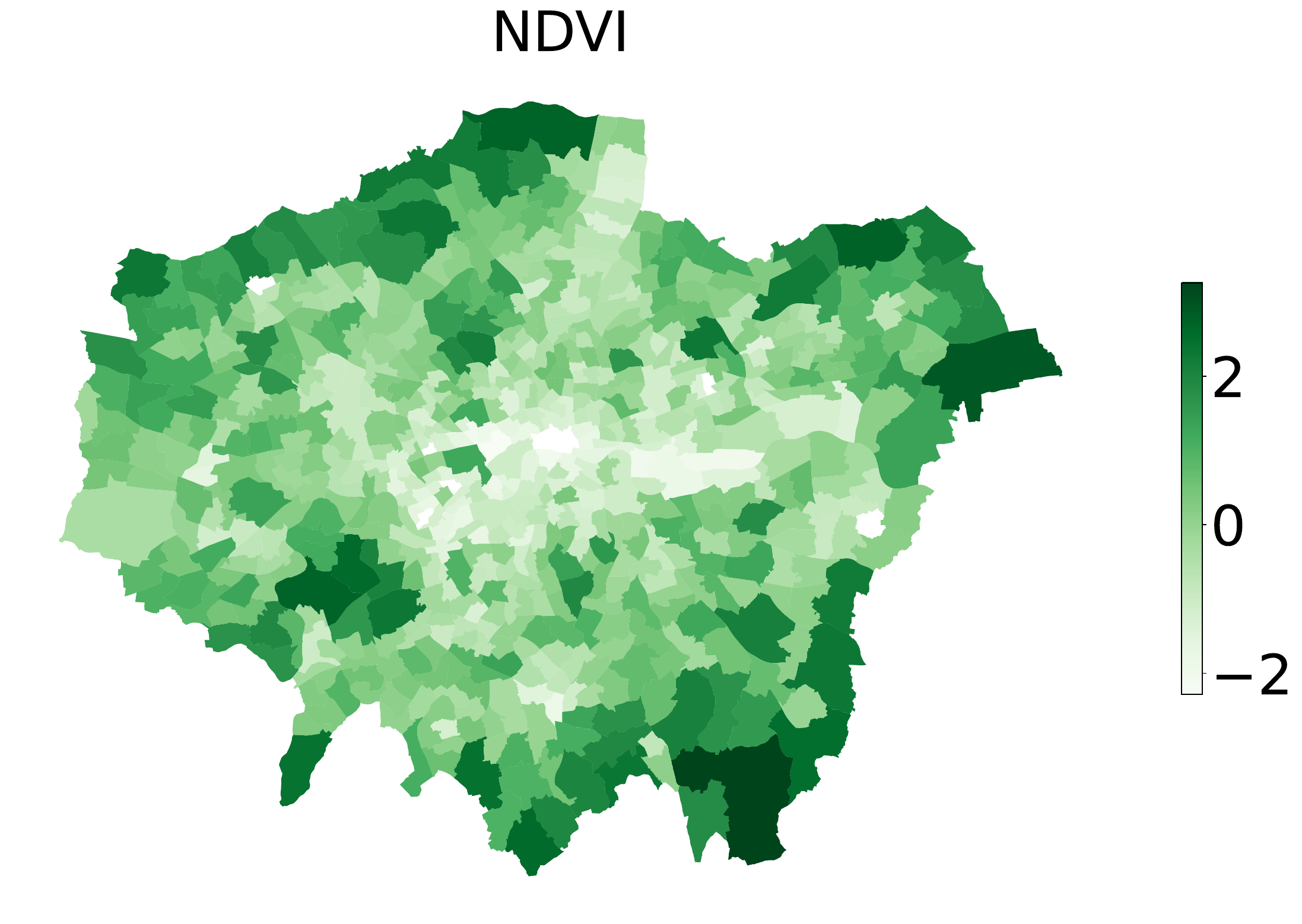}
    \includegraphics[width=.44\textwidth]{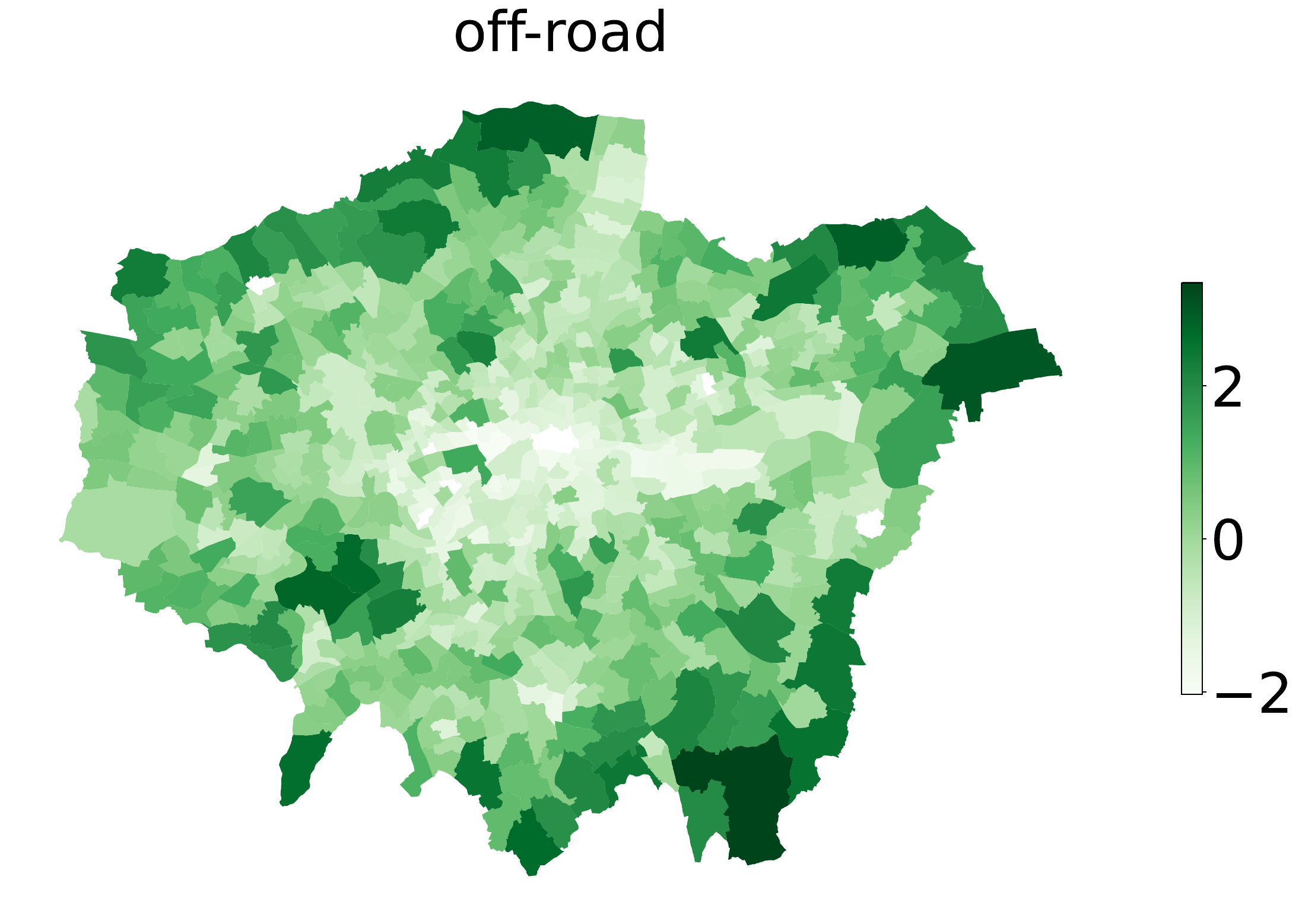}
    \includegraphics[width=.44\textwidth]{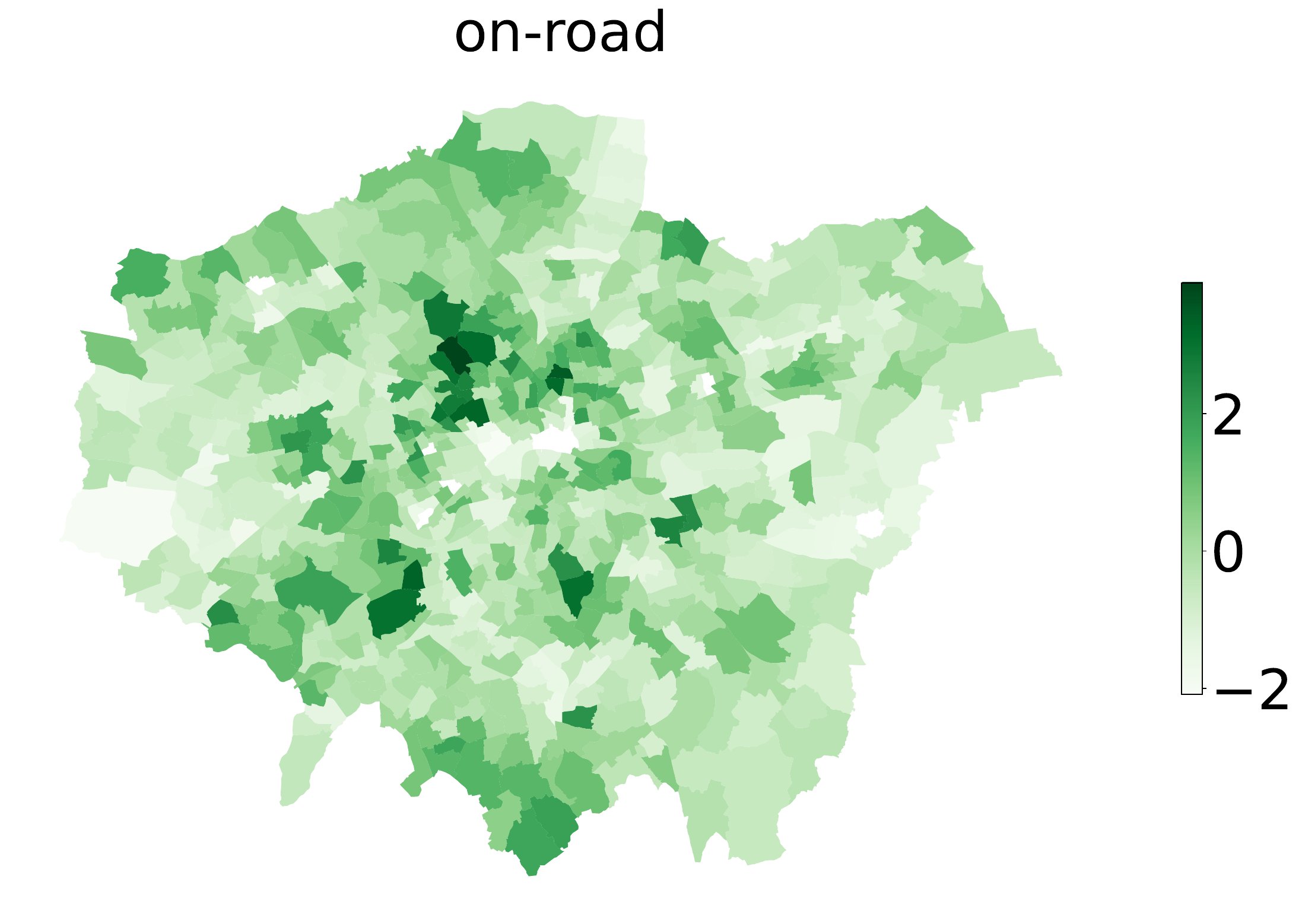}

    \caption{\textbf{Spatial distributions of the average greenery measures across London wards.} Standardized scores are shown, darker colors indicate higher greenery scores.}
    \label{fig_spatial_dist_greenery}
\end{figure*}

\vspace{2em}
The spatial distributions of the averages for the four official and our two proposed greenery measures across London wards are shown in Figure~\ref{fig_spatial_dist_greenery}.

\subsection{Methods for Predicting Prescriptions}
\label{methods_predicting}

\subsubsection{Geographically Weighted Regression (GWR).}  
\label{methods_gwr}
GWR is a spatial modeling technique that extends traditional regression approaches by allowing the relationships between predictors and the dependent variable to vary geographically \cite{brunsdon1998geographically}. Unlike global models such as ordinary least squares or spatial linear model, which assume that these relationships are consistent across space, GWR captures local variations, making it particularly useful for studying spatially non-stationary processes \cite{oshan2019mgwr,brunsdon1998geographically}.  

This method constructs a series of localized regression models at each location by leveraging data from neighboring areas, assigning weights to observations based on their proximity. The GWR equation can be expressed as:  
\[
y_i = \beta_0(u_i, v_i) + \sum_{k=1}^{K} \beta_k(u_i, v_i) x_{ik} + \epsilon_i,
\]  
where \( y_i \) represents the dependent variable at location \(i\);
\( \beta_k(u_i, v_i) \) are the coefficients for the \(k\)-th predictor at geographic coordinates \((u_i, v_i)\);
\( x_{ik} \) is the value of the \(k\)-th predictor at location \(i\); and \( \epsilon_i \) denotes the error term.

GWR enables the investigation of spatial heterogeneity by fitting distinct regression equations for each location. Observations are weighted using a spatial kernel, with closer data points exerting greater influence. For this study, we utilized the \texttt{mgwr} Python library \cite{oshan2019mgwr}, determining that the adaptive bisquare kernel provided the best performance.  

\subsubsection{{PSM Analysis at the Ecological Level}}
\label{methods_causal}
{To move beyond traditional regression analyses, we used propensity score matching (PSM)~\cite{rosenbaum1983central}, a method designed to reduce confounding in observational data and facilitate causal inference, to estimate the average treatment effect (ATE) of greenery on medical prescription outcomes.}

In experimental studies, \emph{randomized control trials} (RCTs) are used to estimate the causal effect of a \emph{treatment} on an \emph{outcome}. RCTs select random subjects, assign a treatment to a subset of them, and finally measure the differences in the outcome between the treated and untreated groups. In observational studies, RCTs are not applicable; instead, \emph{matching} techniques are often used to facilitate inferring causation. Matching works by pairing subjects that were exposed to different either the treatment or outcome but were comparable in terms of \emph{confounding variables}---those factors that may affect being assigned to the treatment or to the control group, or that may affect the outcome. The magnitude of the causal effect is then estimated with the \emph{average treatment effect} (ATE), namely the average difference of the outcome variable between paired subjects:
\begin{equation}
ATE = \frac{\sum_{(s_0,s_1) \in M} y(s_1) - y(s_0)}{|M|},
\label{eqn:ate}
\end{equation}
where $y$ is the outcome, $M$ is the set of paired subjects, and $s_1$ and $s_0$ are two comparable subjects, one ($s1$) in the treatment group, and the other ($s_0$) in the control group.

In our setup, the subjects were London areas (i.e., wards or LSOAs), the treatment was a binary indicator of a given greenery score (\emph{on-road}, \emph{off-road}, \emph{WHO}, \emph{ESA WHO}, \emph{NE}, and \emph{NDVI} greenery) being higher than the median level across the London wards, and the outcome was the min-max normalized value of medical prescription quantity per capita. To match pairs of wards, we used PSM, which matches subjects based on a propensity score, the probability of a subject being assigned to the treatment, given a set of its covariates. We included as confounders the same set of variables as in the regression models, i.e., \emph{IMD score} that captures the area's economic prosperity/deprivation, and population characteristics (\emph{building density}, \emph{median age} and \emph{white percent}).

To ensure robust results and estimate standard errors, we bootstrapped our PSM calculations 1000 times, and we report the average ATEs. 

\paragraph{Significance calculation}
To assess the significance of each ATE, we use the confidence interval method. The 99\% confidence interval is calculated as:

\[
\text{CI}_{99} = \left[ \text{ATE}_{0.5\%}, \text{ATE}_{99.5\%} \right]
\]

where \(\text{ATE}_{0.5\%}\) and \(\text{ATE}_{99.5\%}\) are the 0.5th and 99.5th percentiles of the bootstrapped ATE estimates, respectively. If the confidence interval does not include 0, the ATE is considered statistically significant at the \(p < 0.01\) level.

\paragraph{Calculation of standard errors}
Lastly, the standard errors (SE) were calculated as the standard deviation of the bootstrapped estimates.

\[
SE = \sqrt{\frac{\sum_{i=1}^{B} (\text{ATE}_i - \overline{\text{ATE}})^2}{B-1}}, \quad \overline{\text{ATE}} = \frac{\sum_{i=1}^{B} \text{ATE}_i}{B},
\]
where $B$ is the number of bootstrap samples, $ATE_i$  is the ATE from the $i$-th sample, and $\overline{{ATE}}$ is the mean ATE.

\backmatter

\bigskip

\section*{Acknowledgement}
We would like to express our gratitude to Linus Dietz for his assistance in obtaining and describing the OSM park access data, to Edyta P. Bogucka for her contribution in enhancing the visual presentation of the figures, and to Mark Nieuwenhuijsen for his valuable comments to the manuscript draft. 

\vspace{2em}









\bibliographystyle{plainnat}
\bibliography{sn-bibliography}

\begin{thebibliography}{70}
\providecommand{\natexlab}[1]{#1}
\providecommand{\url}[1]{\texttt{#1}}
\expandafter\ifx\csname urlstyle\endcsname\relax
  \providecommand{\doi}[1]{doi: #1}\else
  \providecommand{\doi}{doi: \begingroup \urlstyle{rm}\Url}\fi

\bibitem[BNF(2019)]{BNFglossary}
{BNF Classifications}.
\newblock
  \url{https://digital.nhs.uk/data-and-information/areas-of-interest/prescribing/practice-level-prescribing-in-england-a-summary/practice-level-prescribing-glossary-of-terms},
  2019.
\newblock [Online; accessed 5-October-2019].

\bibitem[Aerts et~al.(2020)Aerts, Dujardin, Nemery, Van~Nieuwenhuyse,
  Van~Orshoven, Aerts, Somers, Hendrickx, Bruffaerts, Bauwelinck,
  et~al.]{aerts2020residential}
Raf Aerts, Sebastien Dujardin, Benoit Nemery, An~Van~Nieuwenhuyse, Jos
  Van~Orshoven, Jean-Marie Aerts, Ben Somers, Marijke Hendrickx, Nicolas
  Bruffaerts, Mariska Bauwelinck, et~al.
\newblock Residential green space and medication sales for childhood asthma: a
  longitudinal ecological study in belgium.
\newblock \emph{Environmental Research}, 189:\penalty0 109914, 2020.

\bibitem[Alcock et~al.(2017)Alcock, White, Cherrie, Wheeler, Taylor, McInnes,
  Im~Kampe, Vardoulakis, Sarran, Soyiri, et~al.]{alcock2017land}
Ian Alcock, Mathew White, Mark Cherrie, Benedict Wheeler, Jonathon Taylor,
  Rachel McInnes, Eveline~Otte Im~Kampe, Sotiris Vardoulakis, Christophe
  Sarran, Ireneous Soyiri, et~al.
\newblock Land cover and air pollution are associated with asthma
  hospitalisations: A cross-sectional study.
\newblock \emph{Environment international}, 109:\penalty0 29--41, 2017.

\bibitem[Appleton(1996)]{appleton1996experience}
Jay Appleton.
\newblock \emph{The experience of landscape}.
\newblock Wiley Chichester, 1996.

\bibitem[Astell-Burt et~al.(2014)Astell-Burt, Feng, and
  Kolt]{astell2014neighborhood}
Thomas Astell-Burt, Xiaoqi Feng, and Gregory~S Kolt.
\newblock Is neighborhood green space associated with a lower risk of type 2
  diabetes? evidence from 267,072 australians.
\newblock \emph{Diabetes care}, 37\penalty0 (1):\penalty0 197--201, 2014.

\bibitem[Astell-Burt et~al.(2022)Astell-Burt, Navakatikyan, Eckermann, Hackett,
  and Feng]{astell2022urban}
Thomas Astell-Burt, Michael Navakatikyan, Simon Eckermann, Maree Hackett, and
  Xiaoqi Feng.
\newblock Is urban green space associated with lower mental healthcare
  expenditure?
\newblock \emph{Social Science \& Medicine}, 292:\penalty0 114503, 2022.

\bibitem[Authority(2024)]{GreenCover2016}
Greater~London Authority.
\newblock Green cover 2016, 2024.
\newblock URL \url{https://data.london.gov.uk/dataset/green-cover-2024}.
\newblock Data accessed: 2024-07-02.

\bibitem[Badrinarayanan et~al.(2016)Badrinarayanan, Kendall, and
  Cipolla]{badrinarayanan2016segnet}
Vijay Badrinarayanan, Alex Kendall, and Roberto Cipolla.
\newblock Segnet: A deep convolutional encoder-decoder architecture for image
  segmentation, 2016.

\bibitem[Battiston and Schifanella(2024)]{battiston2024need}
Alice Battiston and Rossano Schifanella.
\newblock On the need for a multi-dimensional framework to measure
  accessibility to urban green.
\newblock \emph{npj Urban Sustainability}, 4\penalty0 (1):\penalty0 1--11,
  2024.

\bibitem[Becker et~al.(2022)Becker, Browning, McAnirlin, Yuan, and
  Helbich]{becker2022green}
Douglas~A Becker, Matthew~HEM Browning, Olivia McAnirlin, Shuai Yuan, and Marco
  Helbich.
\newblock Is green space associated with opioid-related mortality? an
  ecological study at the us county level.
\newblock \emph{Urban Forestry \& Urban Greening}, 70:\penalty0 127529, 2022.

\bibitem[Bodicoat et~al.(2014)Bodicoat, O'Donovan, Dalton, Gray, Yates,
  Edwardson, Hill, Webb, Khunti, Davies, et~al.]{bodicoat2014association}
Danielle~H Bodicoat, Gary O'Donovan, Alice~M Dalton, Laura~J Gray, Thomas
  Yates, Charlotte Edwardson, Sian Hill, David~R Webb, Kamlesh Khunti,
  Melanie~J Davies, et~al.
\newblock The association between neighbourhood greenspace and type 2 diabetes
  in a large cross-sectional study.
\newblock \emph{BMJ open}, 4\penalty0 (12):\penalty0 e006076, 2014.

\bibitem[Bratman et~al.(2015)Bratman, Hamilton, Hahn, Daily, and
  Gross]{bratman2015nature}
Gregory~N Bratman, J~Paul Hamilton, Kevin~S Hahn, Gretchen~C Daily, and James~J
  Gross.
\newblock Nature experience reduces rumination and subgenual prefrontal cortex
  activation.
\newblock \emph{Proceedings of the national academy of sciences}, 112\penalty0
  (28):\penalty0 8567--8572, 2015.

\bibitem[Bratman et~al.(2019)Bratman, Anderson, Berman, Cochran, De~Vries,
  Flanders, Folke, Frumkin, Gross, Hartig, et~al.]{bratman2019nature}
Gregory~N Bratman, Christopher~B Anderson, Marc~G Berman, Bobby Cochran, Sjerp
  De~Vries, Jon Flanders, Carl Folke, Howard Frumkin, James~J Gross, Terry
  Hartig, et~al.
\newblock Nature and mental health: An ecosystem service perspective.
\newblock \emph{Science advances}, 5\penalty0 (7):\penalty0 eaax0903, 2019.

\bibitem[Brody et~al.(1987)Brody, Brock, and Williams]{brody1987trends}
Jacob~A Brody, Dwight~B Brock, and T~Franklin Williams.
\newblock Trends in the health of the elderly population.
\newblock \emph{Annual review of public health}, 8\penalty0 (1):\penalty0
  211--234, 1987.

\bibitem[Browning et~al.(2022)Browning, Rigolon, McAnirlin,
  et~al.]{browning2022greenspace}
Matthew~HEM Browning, Alessandro Rigolon, Olivia McAnirlin, et~al.
\newblock Where greenspace matters most: A systematic review of urbanicity,
  greenspace, and physical health.
\newblock \emph{Landscape and Urban Planning}, 217:\penalty0 104233, 2022.

\bibitem[Brunsdon et~al.(1998)Brunsdon, Fotheringham, and
  Charlton]{brunsdon1998geographically}
Chris Brunsdon, Stewart Fotheringham, and Martin Charlton.
\newblock Geographically weighted regression.
\newblock \emph{Journal of the Royal Statistical Society: Series D (The
  Statistician)}, 47\penalty0 (3):\penalty0 431--443, 1998.

\bibitem[Bursik~Jr and Grasmick(1993)]{bursik1993economic}
Robert~J Bursik~Jr and Harold~G Grasmick.
\newblock Economic deprivation and neighborhood crime rates, 1960-1980.
\newblock \emph{Law \& Soc'y Rev.}, 27:\penalty0 263, 1993.

\bibitem[Chang et~al.(2019)Chang, Kim, and Jeon]{chang2019larger}
Yu~Sang Chang, Hann~Earl Kim, and Seongmin Jeon.
\newblock Do larger cities experience lower crime rates? a scaling analysis of
  758 cities in the us.
\newblock \emph{Sustainability}, 11\penalty0 (11):\penalty0 3111, 2019.

\bibitem[Curtis and Goldacre(2018)]{curtis2018openprescribing}
Helen~J Curtis and Ben Goldacre.
\newblock Openprescribing: normalised data and software tool to research trends
  in english nhs primary care prescribing 1998--2016.
\newblock \emph{{BMJ Open}}, 8\penalty0 (2):\penalty0 e019921, 2018.

\bibitem[Curtis et~al.(2019)Curtis, Croker, Walker, Richards, Quinlan, and
  Goldacre]{curtis2019opioid}
Helen~J Curtis, Richard Croker, Alex~J Walker, Georgia~C Richards, Jane
  Quinlan, and Ben Goldacre.
\newblock Opioid prescribing trends and geographical variation in england,
  1998--2018: a retrospective database study.
\newblock \emph{The Lancet Psychiatry}, 6\penalty0 (2):\penalty0 140--150,
  2019.

\bibitem[Dendup et~al.(2018)Dendup, Feng, Clingan, and
  Astell-Burt]{dendup2018environmental}
Tashi Dendup, Xiaoqi Feng, Stephanie Clingan, and Thomas Astell-Burt.
\newblock Environmental risk factors for developing type 2 diabetes mellitus: a
  systematic review.
\newblock \emph{International journal of environmental research and public
  health}, 15\penalty0 (1):\penalty0 78, 2018.

\bibitem[Drukker et~al.(2003)Drukker, Kaplan, Feron, and
  Van~Os]{drukker2003children}
Marjan Drukker, Charles Kaplan, Frans Feron, and Jim Van~Os.
\newblock Children's health-related quality of life, neighbourhood
  socio-economic deprivation and social capital. a contextual analysis.
\newblock \emph{Social science \& medicine}, 57\penalty0 (5):\penalty0
  825--841, 2003.

\bibitem[Eisenman et~al.(2019)Eisenman, Churkina, Jariwala, Kumar, Lovasi,
  Pataki, Weinberger, and Whitlow]{eisenman2019urban}
Theodore~S Eisenman, Galina Churkina, Sunit~P Jariwala, Prashant Kumar, Gina~S
  Lovasi, Diane~E Pataki, Kate~R Weinberger, and Thomas~H Whitlow.
\newblock Urban trees, air quality, and asthma: An interdisciplinary review.
\newblock \emph{Landscape and urban planning}, 187:\penalty0 47--59, 2019.

\bibitem[Fotheringham and Wong(1991)]{fotheringham1991modifiable}
A~Stewart Fotheringham and David~WS Wong.
\newblock The modifiable areal unit problem in multivariate statistical
  analysis.
\newblock \emph{Environment and planning A}, 23\penalty0 (7):\penalty0
  1025--1044, 1991.

\bibitem[Gil et~al.(2015)Gil, Varoudis, Karimi, and Penn]{gil2015space}
Jorge Gil, Tasos Varoudis, Kayvan Karimi, and Alan Penn.
\newblock The space syntax toolkit: Integrating depthmapx and exploratory
  spatial analysis workflows in qgis.
\newblock In \emph{SSS 2015-10th International Space Syntax Symposium},
  volume~10. Space Syntax Laboratory, The Bartlett School of Architecture,
  UCL~…, 2015.

\bibitem[Hillier(2007)]{hillier2007space}
Bill Hillier.
\newblock \emph{Space is the machine: a configurational theory of
  architecture}.
\newblock Space Syntax, 2007.

\bibitem[Hillier and Hanson(1989)]{hillier1989social}
Bill Hillier and Julienne Hanson.
\newblock \emph{The social logic of space}.
\newblock Cambridge university press, 1989.

\bibitem[Hillier and Iida(2005)]{hillier2005network}
Bill Hillier and Shinichi Iida.
\newblock Network and psychological effects in urban movement.
\newblock In \emph{International Conference on Spatial Information Theory},
  pages 475--490. Springer, 2005.

\bibitem[Hyam(2020)]{hyam2020greenness}
Roger Hyam.
\newblock Greenness, mortality and mental health prescription rates in urban
  scotland-a population level, observational study.
\newblock \emph{Research Ideas and Outcomes}, 6:\penalty0 e53542, 2020.

\bibitem[Intelligence(2019)]{greenlondond}
GLA~City Intelligence.
\newblock How green is london?
\newblock Technical report, GLAIntelligence, 2019.

\bibitem[Jiang et~al.(2018)Jiang, Luo, Xu, and Wang]{jiang2018does}
J~Jiang, L~Luo, P~Xu, and P~Wang.
\newblock How does social development influence life expectancy? a
  geographically weighted regression analysis in china.
\newblock \emph{Public health}, 163:\penalty0 95--104, 2018.

\bibitem[Joglekar et~al.(2020)Joglekar, Quercia, Redi, Aiello, Kauer, and
  Sastry]{joglekar2020facelift}
Sagar Joglekar, Daniele Quercia, Miriam Redi, Luca~Maria Aiello, Tobias Kauer,
  and Nishanth Sastry.
\newblock Facelift: a transparent deep learning framework to beautify urban
  scenes.
\newblock \emph{Royal Society open science}, 7\penalty0 (1):\penalty0 190987,
  2020.

\bibitem[Kaplan(1995)]{kaplan1995restorative}
Stephen Kaplan.
\newblock The restorative benefits of nature: Toward an integrative framework.
\newblock \emph{Journal of environmental psychology}, 15\penalty0 (3):\penalty0
  169--182, 1995.

\bibitem[Kellert and Wilson(1995)]{biophiliahypothesis}
S.R. Kellert and E.O. Wilson.
\newblock The biophilia hypothesis.
\newblock In \emph{Island Press}, 1995.

\bibitem[Kirkman et~al.(2012)Kirkman, Briscoe, Clark, Florez, Haas, Halter,
  Huang, Korytkowski, Munshi, Odegard, et~al.]{kirkman2012diabetes}
M~Sue Kirkman, Vanessa~Jones Briscoe, Nathaniel Clark, Hermes Florez, Linda~B
  Haas, Jeffrey~B Halter, Elbert~S Huang, Mary~T Korytkowski, Medha~N Munshi,
  Peggy~Soule Odegard, et~al.
\newblock Diabetes in older adults.
\newblock \emph{Diabetes care}, 35\penalty0 (12):\penalty0 2650--2664, 2012.

\bibitem[Knox et~al.(2010)Knox, Law, Jewison, Liu, Ly, Frolkis, Pon, Banco,
  Mak, Neveu, et~al.]{knox2010drugbank}
Craig Knox, Vivian Law, Timothy Jewison, Philip Liu, Son Ly, Alex Frolkis,
  Allison Pon, Kelly Banco, Christine Mak, Vanessa Neveu, et~al.
\newblock Drugbank 3.0: a comprehensive resource for `omics' research on drugs.
\newblock \emph{Nucleic acids research}, 39\penalty0 (suppl\_1):\penalty0
  D1035--D1041, 2010.

\bibitem[Law et~al.(2018)Law, Seresinhe, Shen, and Gutierrez-Roig]{Law2018}
Stephen Law, Chanuki~Illushka Seresinhe, Yao Shen, and Mario Gutierrez-Roig.
\newblock Street-frontage-net: urban image classification using deep
  convolutional neural networks.
\newblock \emph{International Journal of Geographical Information Science},
  0\penalty0 (0):\penalty0 1--27, 2018.
\newblock \doi{10.1080/13658816.2018.1555832}.
\newblock URL \url{https://doi.org/10.1080/13658816.2018.1555832}.

\bibitem[Law et~al.(2019)Law, Paige, and Russell]{law2019take}
Stephen Law, Brooks Paige, and Chris Russell.
\newblock Take a look around: using street view and satellite images to
  estimate house prices.
\newblock \emph{ACM Transactions on Intelligent Systems and Technology (TIST)},
  10\penalty0 (5):\penalty0 1--19, 2019.

\bibitem[Leng et~al.(2020)Leng, Li, Yan, and An]{leng2020exploring}
Hong Leng, Shuyuan Li, Shichun Yan, and Xiuli An.
\newblock Exploring the relationship between green space in a neighbourhood and
  cardiovascular health in the winter city of china: a study using a health
  survey for harbin.
\newblock \emph{International Journal of Environmental Research and Public
  Health}, 17\penalty0 (2):\penalty0 513, 2020.

\bibitem[Li et~al.(2022)Li, Zhang, Hu, He, Yang, Zhao, Zhu, Zhu, and
  Huang]{li2022associations}
Guoao Li, Hanshuang Zhang, Mingjun Hu, Jialiu He, Wanjun Yang, Huanhuan Zhao,
  Zhenyu Zhu, Jinliang Zhu, and Fen Huang.
\newblock Associations of combined exposures to ambient temperature, air
  pollution, and green space with hypertension in rural areas of anhui
  province, china: A cross-sectional study.
\newblock \emph{Environmental Research}, 204:\penalty0 112370, 2022.

\bibitem[Luxen and Vetter(2011)]{luxen-vetter-2011}
Dennis Luxen and Christian Vetter.
\newblock Real-time routing with openstreetmap data.
\newblock In \emph{Proceedings of the 19th ACM SIGSPATIAL International
  Conference on Advances in Geographic Information Systems}, GIS '11, pages
  513--516, New York, NY, USA, 2011. ACM.
\newblock ISBN 978-1-4503-1031-4.
\newblock \doi{10.1145/2093973.2094062}.

\bibitem[MacKerron and Mourato(2013)]{MACKERRON2013992}
George MacKerron and Susana Mourato.
\newblock Happiness is greater in natural environments.
\newblock \emph{Global Environmental Change}, 23\penalty0 (5):\penalty0 992 --
  1000, 2013.
\newblock ISSN 0959-3780.
\newblock \doi{https://doi.org/10.1016/j.gloenvcha.2013.03.010}.
\newblock URL
  \url{http://www.sciencedirect.com/science/article/pii/S0959378013000575}.

\bibitem[Map(2023)]{OSM_overpass}
Open~Street Map.
\newblock {Overpass API}.
\newblock \url{https://overpass-turbo.eu}, 2023.
\newblock [Online; accessed 22-March-2023].

\bibitem[Mazumdar et~al.(2021)Mazumdar, Chong, Astell-Burt, Feng, Morgan, and
  Jalaludin]{mazumdar2021green}
Soumya Mazumdar, Shanley Chong, Thomas Astell-Burt, Xiaoqi Feng, Geoffrey
  Morgan, and Bin Jalaludin.
\newblock Which green space metric best predicts a lowered odds of type 2
  diabetes?
\newblock \emph{International Journal of Environmental Research and Public
  Health}, 18\penalty0 (8):\penalty0 4088, 2021.

\bibitem[Mears et~al.(2021)Mears, Brindley, Barrows, Richardson, and
  Maheswaran]{mears2021mapping}
Meghann Mears, Paul Brindley, Paul Barrows, Miles Richardson, and Ravi
  Maheswaran.
\newblock Mapping urban greenspace use from mobile phone gps data.
\newblock \emph{Plos one}, 16\penalty0 (7):\penalty0 e0248622, 2021.

\bibitem[{Natural England}(2010)]{england2010nature}
{Natural England}.
\newblock Nature nearby: accessible natural greenspace guidance.
\newblock
  \url{http://www.ukmaburbanforum.co.uk/docunents/other/nature_nearby.pdf},
  2010.

\bibitem[Ng et~al.(2010)Ng, Agarwal, Sidiki, McKay, Townend, and
  Azuara-Blanco]{ng2010effect}
Wai~Siene Ng, Pankaj~Kumar Agarwal, Sikander Sidiki, L~McKay, John Townend, and
  Augusto Azuara-Blanco.
\newblock The effect of socio-economic deprivation on severity of glaucoma at
  presentation.
\newblock \emph{British journal of ophthalmology}, 94\penalty0 (1):\penalty0
  85--87, 2010.

\bibitem[{NHS Business Services Authority (NHSBSA)}()]{NHSBSA_EPD_Resource}
{NHS Business Services Authority (NHSBSA)}.
\newblock {English Prescribing Dataset (EPD)}.
\newblock URL
  \url{https://opendata.nhsbsa.net/dataset/english-prescribing-data-epd/resource/11d9e694-69d7-4ed5-bf3d-6035a4ba02f9}.
\newblock Data resource within the English Prescribing Data (EPD) dataset.
  Contains detailed information on prescriptions issued in England and
  dispensed within the UK.

\bibitem[{NHS Digital}(2019)]{PracticeLevelPrescribing}
{NHS Digital}.
\newblock {BNF Classifications for Practice-Level Prescribing Data}, 2019.
\newblock URL
  \url{https://digital.nhs.uk/data-and-information/areas-of-interest/prescribing/practice-level-prescribing-in-england-a-summary}.

\bibitem[{OpenStreetMap contributors}(2017)]{OpenStreetMap}
{OpenStreetMap contributors}.
\newblock {Planet dump retrieved from https://planet.osm.org }.
\newblock \url{ https://www.openstreetmap.org }, 2017.

\bibitem[Oshan et~al.(2019)Oshan, Li, Kang, Wolf, and
  Fotheringham]{oshan2019mgwr}
Taylor~M Oshan, Ziqi Li, Wei Kang, Levi~J Wolf, and A~Stewart Fotheringham.
\newblock mgwr: A python implementation of multiscale geographically weighted
  regression for investigating process spatial heterogeneity and scale.
\newblock \emph{ISPRS International Journal of Geo-Information}, 8\penalty0
  (6):\penalty0 269, 2019.

\bibitem[Penn et~al.(1998)Penn, Hillier, Banister, and
  Xu]{penn1998configurational}
Alan Penn, Bill Hillier, David Banister, and Jun Xu.
\newblock Configurational modelling of urban movement networks.
\newblock \emph{Environment and Planning B: planning and design}, 25\penalty0
  (1):\penalty0 59--84, 1998.

\bibitem[Pettorelli(2013)]{pettorelli2013normalized}
Nathalie Pettorelli.
\newblock \emph{The normalized difference vegetation index}.
\newblock Oxford University Press, USA, 2013.

\bibitem[Rosenbaum and Rubin(1983)]{rosenbaum1983central}
Paul~R Rosenbaum and Donald~B Rubin.
\newblock The central role of the propensity score in observational studies for
  causal effects.
\newblock \emph{Biometrika}, 70\penalty0 (1):\penalty0 41--55, 1983.

\bibitem[Rust{\o}en et~al.(2005)Rust{\o}en, Wahl, Hanestad, Lerdal, Paul, and
  Miaskowski]{rustoen2005age}
Tone Rust{\o}en, Astrid~Klopstad Wahl, Berit~Rokne Hanestad, Anners Lerdal,
  Steven Paul, and Christine Miaskowski.
\newblock Age and the experience of chronic pain: differences in health and
  quality of life among younger, middle-aged, and older adults.
\newblock \emph{The Clinical journal of pain}, 21\penalty0 (6):\penalty0
  513--523, 2005.

\bibitem[Scepanovic et~al.(2024)Scepanovic, Obadic, Joglekar, Giustarini,
  Nattero, Quercia, and Zhu]{scepanovic2024medsat}
Sanja Scepanovic, Ivica Obadic, Sagar Joglekar, Laura Giustarini, Cristiano
  Nattero, Daniele Quercia, and Xiaoxiang Zhu.
\newblock Medsat: A public health dataset for england featuring medical
  prescriptions and satellite imagery.
\newblock \emph{Advances in Neural Information Processing Systems}, 36, 2024.

\bibitem[{Schiavina, M. Freire, S. MacManus, K.}(2023)]{ghs_pop}
{Schiavina, M. Freire, S. MacManus, K.}
\newblock Ghs population grid multitemporal (1975-2030) r2023a.
\newblock European Commission, Joint Research Centre (JRC) [Dataset], 2023.

\bibitem[Schifanella et~al.(2020)Schifanella, Vedove, Salomone, Bajardi, and
  Paolotti]{schifanella2020spatial}
Rossano Schifanella, Dario~Delle Vedove, Alberto Salomone, Paolo Bajardi, and
  Daniela Paolotti.
\newblock Spatial heterogeneity and socioeconomic determinants of opioid
  prescribing in england between 2015 and 2018.
\newblock \emph{BMC medicine}, 18:\penalty0 1--13, 2020.

\bibitem[Seresinhe et~al.(2015)Seresinhe, Preis, and Moat]{Seresinhe2015}
Chanuki~Illushka Seresinhe, Tobias Preis, and Helen~Susannah Moat.
\newblock Quantifying the impact of scenic environments on health.
\newblock In \emph{Nature Scientific Reports}, 2015.

\bibitem[Shepley et~al.(2019)Shepley, Sachs, Sadatsafavi, Fournier, and
  Peditto]{shepley2019impact}
Mardelle Shepley, Naomi Sachs, Hessam Sadatsafavi, Christine Fournier, and Kati
  Peditto.
\newblock The impact of green space on violent crime in urban environments: an
  evidence synthesis.
\newblock \emph{International journal of environmental research and public
  health}, 16\penalty0 (24):\penalty0 5119, 2019.

\bibitem[S{\o}rensen et~al.(2022)S{\o}rensen, Poulsen, Hvidtfeldt, Brandt,
  Frohn, Ketzel, Christensen, Im, Khan, M{\"u}nzel, et~al.]{sorensen2022air}
Mette S{\o}rensen, Aslak~H Poulsen, Ulla~A Hvidtfeldt, J{\o}rgen Brandt, Lise~M
  Frohn, Matthias Ketzel, Jesper~H Christensen, Ulas Im, Jibran Khan, Thomas
  M{\"u}nzel, et~al.
\newblock Air pollution, road traffic noise and lack of greenness and risk of
  type 2 diabetes: a multi-exposure prospective study covering denmark.
\newblock \emph{Environment International}, 170:\penalty0 107570, 2022.

\bibitem[Su et~al.(2017)Su, Gong, Tan, Pi, Weng, and Cai]{su2017area}
Shiliang Su, Yue Gong, Bingqing Tan, Jianhua Pi, Min Weng, and Zhongliang Cai.
\newblock Area social deprivation and public health: Analyzing the spatial
  non-stationary associations using geographically weighed regression.
\newblock \emph{Social Indicators Research}, 133:\penalty0 819--832, 2017.

\bibitem[Sun et~al.(2011)Sun, Chu, and Sung]{sun2011cross}
Ivan~Y Sun, Doris~C Chu, and Hung-En Sung.
\newblock A cross-national analysis of the mediating effect of economic
  deprivation on crime.
\newblock \emph{Asian Journal of Criminology}, 6\penalty0 (1):\penalty0 15--32,
  2011.

\bibitem[Turunen et~al.(2023)Turunen, Halonen, Korpela, Ojala, Pasanen,
  Siponen, Tiittanen, Tyrv{\"a}inen, Yli-Tuomi, and Lanki]{turunen2023cross}
Anu~W Turunen, Jaana Halonen, Kalevi Korpela, Ann Ojala, Tytti Pasanen, Taina
  Siponen, Pekka Tiittanen, Liisa Tyrv{\"a}inen, Tarja Yli-Tuomi, and Timo
  Lanki.
\newblock Cross-sectional associations of different types of nature exposure
  with psychotropic, antihypertensive and asthma medication.
\newblock \emph{Occupational and environmental medicine}, 80\penalty0
  (2):\penalty0 111--118, 2023.

\bibitem[Varoudis et~al.(2013)Varoudis, Law, Karimi, Hillier, and
  Penn]{varoudis2013space}
Tasos Varoudis, Stephen Law, Kayvan Karimi, Bill Hillier, and Alan Penn.
\newblock Space syntax angular betweenness centrality revisited.
\newblock In \emph{Proceedings of 9th International Space Syntax Symposium,
  Seoul}, 2013.

\bibitem[Ye et~al.(2019{\natexlab{a}})Ye, Richards, Lu, Song, Zhuang, Zeng, and
  Zhong]{ye2019measuring}
Yu~Ye, Daniel Richards, Yi~Lu, Xiaoping Song, Yu~Zhuang, Wei Zeng, and Teng
  Zhong.
\newblock Measuring daily accessed street greenery: A human-scale approach for
  informing better urban planning practices.
\newblock \emph{Landscape and Urban Planning}, 191:\penalty0 103434,
  2019{\natexlab{a}}.

\bibitem[Ye et~al.(2019{\natexlab{b}})Ye, Xie, Fang, Jiang, and
  Wang]{ye2019daily}
Yu~Ye, Hanting Xie, Jia Fang, Hetao Jiang, and De~Wang.
\newblock Daily accessed street greenery and housing price: Measuring economic
  performance of human-scale streetscapes via new urban data.
\newblock \emph{Sustainability}, 11\penalty0 (6):\penalty0 1741,
  2019{\natexlab{b}}.

\bibitem[Yu and Kwan(2024)]{yu2024dynamic}
Changda Yu and Mei-Po Kwan.
\newblock Dynamic greenspace exposure, individual mental health status and
  momentary stress level: A study using multiple greenspace measurements.
\newblock \emph{Health \& Place}, 86:\penalty0 103213, 2024.

\bibitem[{Zanaga, D. et al}(2021)]{zanaga_daniele_2021_5571936}
{Zanaga, D. et al}.
\newblock Esa worldcover 10 m 2020 v100 [dataset].
\newblock 10.5281/zenodo.5571936, 2021.

\bibitem[Zierler et~al.(2000)Zierler, Krieger, Tang, Coady, Siegfried, DeMaria,
  and Auerbach]{zierler2000economic}
Sally Zierler, Nancy Krieger, Yuren Tang, William Coady, Erika Siegfried,
  Alfred DeMaria, and John Auerbach.
\newblock Economic deprivation and aids incidence in massachusetts.
\newblock \emph{American journal of public health}, 90\penalty0 (7):\penalty0
  1064, 2000.

\end{thebibliography}


\end{document}